\documentclass[showpacs, amsmath, amssymb, onecolumn, notitlepage]{revtex4-1}

\usepackage{appendix}
\usepackage{bbm}
\usepackage{braket}
\usepackage{amsfonts}
\usepackage{amsthm}
\usepackage{amsmath}

\usepackage{txfonts}
\usepackage{bm}
\usepackage[dvipdfmx]{graphicx}
\usepackage{color}
\usepackage{mathrsfs}
\usepackage{subfigure}
\usepackage{wrapfig}
\usepackage{here}
\usepackage{hyperref}
\usepackage{stackengine}

\newcommand{\be}{\begin{equation}}
\newcommand{\ee}{\end{equation}}
\newcommand{\ba}{\begin{align}}
\newcommand{\ea}{\end{align}}

\newcommand{\barbelow}[1]{\stackunder[1.2pt]{$#1$}{\rule{.8ex}{.14ex}}}

\begin{document}

\title{Security of the BB84 protocol with passive biased basis choice by the receiver}






\author{
Shun Kawakami$^{1,2}$, Atsushi Taniguchi$^{1}$, Yoshihide Tonomura$^{1}$, Koichi Takasugi$^{1}$ and Koji Azuma$^{3,2}$}

\affiliation{
$^1$ Network Innovation Laboratories, NTT Inc.\\ 1-1 Hikari-no-oka, Yokosuka, Kanagawa, 239-0847, Japan \\
$^2$ NTT Research Center for Theoretical Quantum Information, NTT Inc.\\ \mbox{3-1 Morinosato Wakamiya, Atsugi, Kanagawa,~243-0198, Japan}\\
$^3$ Basic Research Laboratories, NTT Inc.\\ \mbox{3-1 Morinosato Wakamiya, Atsugi, Kanagawa,~243-0198, Japan}\\
}
\begin{abstract}
The Bennett-Brassard 1984 protocol (BB84 protocol) is one of the simplest protocols for implementing quantum key distribution (QKD). 
In the protocol, the sender and the receiver iteratively choose one of two complementary measurement bases. 
Regarding the basis choice by the receiver, a passive setup has been adopted in a number of its implementations including satellite QKD and time-bin encoding one. 
However, conventional theoretical techniques to prove the security of BB84 protocol are not applicable if the receiver chooses his measurement basis passively, rather than actively,  with a biased probability, followed by the measurement with threshold detectors. 
Here we present a fully analytical security proof against coherent attacks for such a decoy-state BB84 protocol with receiver's passive basis choice and measurement with threshold detectors. 
The numerical simulations under practical situations show that the difference in secure key rate between the active and the passive implementations of the protocol is negligible except for long communication distances. \end{abstract}

\maketitle

\section{Introduction}
Quantum key distribution (QKD) is an established research field among a number of quantum information processing technologies from the perspective of both theory and implementation. 
Many QKD protocols have been proposed so far, but the Bennett-Brassard 1984 protocol (BB84 protocol) \cite{1984Bennett} is one of most frequently demonstrated protocols and its security analysis is studied well. 
Although the sender is assumed to emit a single photon in the ideal BB84 protocol, in practice weak coherent pulses can be used instead with decoy-state method \cite{2003Hwang,  2005WangPRL, 2005Lo, 2005Wang}, in which the pulse intensities are modulated, to estimate the contribution of single-photon emission to key generation. 
Regarding the receiver, although a single photon is assumed to be received in the ideal BB84 protocol, arrival of single photon cannot be distinguished from that of multiple photons because threshold (or on/off) detectors are used in practice. 
Nevertheless, the security of such a decoy-state BB84 protocol implemented with threshold detectors is proven thanks to sophisticated theoretical techniques such as 
 the squashing method \cite{2008Lutkenhaus, 2008Tsurumaru} and security proofs with complementarity \cite{2006Koashi, 2009Koashi} and with entropic uncertainty relation \cite{2011Tomamichel}. 

In the conventional practical BB84 protocol, the sender and the receiver actively choose one of two complementary bases, called $Z$ basis and $X$ basis, with a probability, perhaps with a biased one to achieve a higher key rate \cite{2004Lo}. We call it such a biased one `active-biased BB84' protocol. 
On the other hand, from the perspective of implementation, a passive basis choice by the receiver, in which a single beam splitter is set to split the incoming light into two passes corresponding to the bases, is a realistic option 
especially in the satellite-based QKD with polarization encoding \cite{2017Liao, 2022Sivasankaran, 2023Shields, 2025Yang} and the time-bin phase encoding BB84 protocol  \cite{2020Yin, 2022Scalon, 2023Tang, 2023Grunenfelder, 2024Francesconi}. 
For this kind of receiver's passive setup, a basis choice with an unbalanced probability is worth being considered in order to not only achieve a higher key generation rate but also relax the required absolute precision in the splitting ratio of 50:50 of the beam splitter used for unbiased basis choice. 
However, the above theoretical techniques for the BB84 protocol with threshold detectors cannot directly be applied to such a `passive-biased BB84' protocol. 
For the passive-biased BB84 protocol, the squashing map does not exist \cite{2024Kamin}. 
Furthermore, the security proof with complementarity as well as with entropic uncertainty relation can be applied to the passive protocol only if the basis choice is balanced \cite{2025Tupkary} (as explicitly shown below with Appendix \ref{Difference}).

Recently, security of the decoy-state QKD protocol with passive-biased basis choice is proven \cite{2024Kamin} by using a Flag-state squasher \cite{2021Zhang}, in which an infinite photon-number space is analytically squashed to a finite photon-number space so as to numerically solve an optimization problem in the finite space. However, their proof has been based on `IID assumption', i.e., the quantum states shared between the sender and the receiver after the interference by an eavesdropper have been assumed to be identical and independent among all rounds of the protocol, until very recently \cite{2025KaminPost, 2025KaminRenyi}. 
Besides, compared with conventional analytic approaches \cite{2006Koashi, 2009Koashi, 2011Tomamichel}, this type of numerical approach necessitates an additional workload to calculate the amount of privacy amplification needed from the observed data in the quantum phase of the protocol, before entering the classical post-processing phase. 



In this paper, in contrast, we present a fully analytical security proof of the decoy-state BB84 protocol with receiver's passive-biased basis choice against coherent attacks in the asymptotic regime. This is done by applying the security proof for BB84 protocol with complementarity \cite{2006Koashi, 2009Koashi}, which inherently covers coherent attacks, {\it only} to the single-photon subspace at the receiver. Our proof explicitly gives the key generation rate including the effect of dark counts and asymmetric loss between two bases. Our work shows that the difference in key generation rate between the active-biased and the passive-biased BB84 protocols is negligible except for the range of long distances where the dark count rate of detectors is relevant to the key rate. This implies that the passive setup of the receiver does not sacrifice the efficiency in key generation despite its implementational benefits. 

This paper is organized as follows. In Section \ref{protocolassumption}, we describe the BB84 protocol with receiver's passive basis choice and assumptions on apparatuses. 
Section \ref{Proof} gives the security proof of the protocol, and shows a formula for the key rate. In Section \ref{numerical analysis}, numerical results are shown and compared with the active protocol. Section \ref{conclusion} gives conclusion.


\section{Protocol and assumptions} \label{protocolassumption}
Here we introduce a polarization-type BB84 protocol with the decoy-state method where the receiver selects his basis passively. The sender is called Alice and the receiver is called Bob.  
In the protocol, two complementary bases ($Z$ and $X$) are used to generate non-orthogonal states where the secret key is extracted from $Z$ basis and the signal disturbance is monitored in $X$ basis. 
The protocol is described as follows:\\

\begin{figure}[t]
 \centering
 \includegraphics[keepaspectratio, scale=0.6]
      {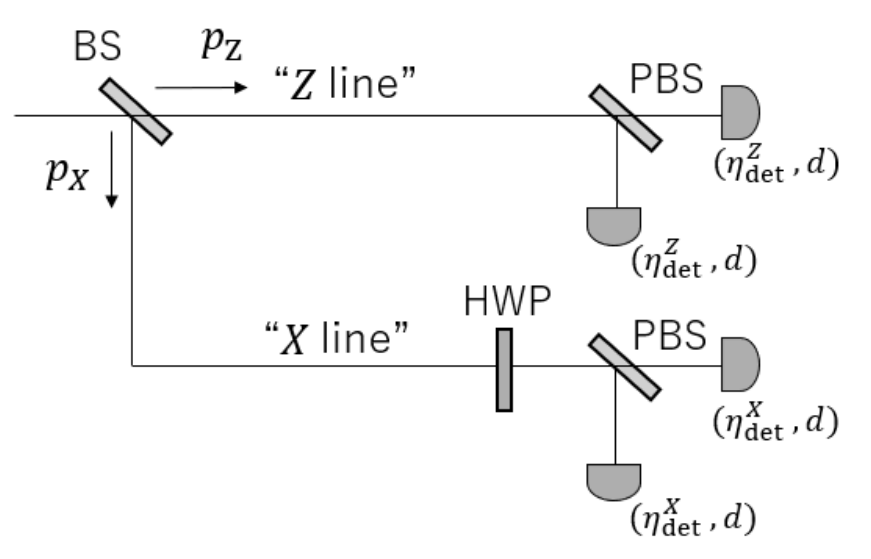}
 \caption{Bob's actual setup for passive basis choice with a beam splitter (BS), polarization beam splitters (PBSs), a half wave plate (HWP) and threshold detectors. 
  Incoming light is split into $Z$ line or $X$ line with the ratio of $p_Z$ to $p_X$. All detectors have the identical dark count probability $d$. The quantum efficiencies of detectors in $Z$ line and $X$ line are $\eta_{\rm det}^Z$ and $\eta_{\rm det}^X$, respectively. The overall transmittances, including the quantum efficiency, in $Z$ line and $X$ line are $\eta_Z$ and $\eta_X$, respectively ($\eta_Z \geq \eta_X$).}
 \label{setup1}
\end{figure}

\noindent
(1) {\it State Preparation}: 
Alice chooses her bit $b_A \in \{0,1\}$ at random and basis $W_A\in \{Z, X\}$ with probability $p_{W_A}$ ($p_Z \geq p_X$, $p_Z + p_X =1$) to determine a polarization $P \in \{H, V, D, \bar{D}\}$, where $P=H, V, D$ and $\bar{D}$ are chosen when $(W_A, b_A) = (Z,0), (Z,1),  (X,0)$ and $(X,1)$, respectively.  She chooses intensity $\mu_A \in \{\mu, \nu, 0\}$ ($\mu > \nu > 0$) with probability $p_{\mu_A}$ $(p_{\mu}+p_{\nu}+p_{0}=1)$. She also chooses phase $\delta_A \in [0, 2\pi )$ at random, which will be never revealed. 
Accordingly, she prepares a perfect coherent state
\be
\ket{\alpha_{P}}_{A'} \coloneqq e^{-\mu_A/2} \sum\limits_{n_A\geq0} \frac{(\sqrt{\mu_A} e^{i \delta_A} \hat{a}_{P}^{\dagger})^{n_A}}{n_A!} \ket{\rm vac}_{A'},
\ee
where $\ket{\rm vac}_{A'}$ represents the vacuum state and $\hat{a}_{P}^{\dagger}$ is the creation operator of polarization $P \in \{H, V, D, \bar{D}\}$ satisfying 
$\hat{a}_{D}^{\dagger} = (\hat{a}_{H}^{\dagger} + \hat{a}_{V}^{\dagger})/\sqrt{2}$ and $\hat{a}_{\bar{D}}^{\dagger} = (\hat{a}_{H}^{\dagger} - \hat{a}_{V}^{\dagger})/\sqrt{2}$. 
She sends the coherent state to Bob. 
Since $\delta_A$ is chosen at random and is never revealed, her state is regarded as a mixture of Fock states; 
\begin{align}
\frac{1}{2\pi} \int_0^{2\pi} \ket{\alpha_{P}}_{A'} \bra{\alpha_{P}} d\delta_A  = 
\sum\limits_{n_A\geq0} P_{\mu_A}^{\rm Poisson}(n_A) \ket{P^{(n_A)}}_{A'} \bra{P^{(n_A)}},
\end{align}
where $P_{\mu_A}^{\rm Poisson}(n_A) \coloneqq e^{-\mu_A}\mu_A^{n_A} /n_A!~(\mu_A \in \{ \mu, \nu, 0 \})$ is a Poisson distribution and 
$\ket{P^{(n_A)}}_{A'} \coloneqq (\hat{a}_{P}^{\dagger})^{n_A} / \sqrt{n_A !}  \ket{\rm vac}_{A'}$ is a $n_A$-photon-number state with a polarization $P \in \{ H,V,D,\bar{D}\}$. 
\\
(2) {\it Measurement}: 
With the linear optical setup in Fig.~\ref{setup1}, Bob passively splits an incoming signal to `$Z$ line' and `$X$ line' to perform $Z$-basis measurement and $X$-basis measurement, respectively, with splitting ratio $p_Z$ to $p_X$ by using a beam splitter (BS). In $X$ line, a half wave plate (HWP), whose transmittance might be smaller than unity, is set to rotate the polarization by 45 degrees. In both lines, a polarization beam splitter (PBS) splits the pass depending on whether the polarization is horizontal or vertical. 
Four threshold detectors have a dark count probability $0 \leq d \leq1$ per pulse. 
In $Z$ ($X$) line, two detectors have an identical quantum efficiency $\eta_{\rm det}^Z$ ($\eta_{\rm det}^X$) and the overall transmittance of the line, including the quantum efficiency, is $\eta_Z$ ($\eta_X$),  satisfying $0 \leq \eta_X \leq \eta_Z \leq 1$ (as $X$ line may have additional optical components such as a half wave plate). Define a parameter  
\be
r\coloneqq \frac{\eta_X}{\eta_Z}~(\leq 1), \label{rrr}
\ee
representing the asymmetric transmittance between the two lines. 
He registers $y_B=1$ if at least one of four detectors clicks and $y_B=0$ if no detector clicks. 
If $y_B=1$ and no detection occurs in $X$ $(Z)$ line, he regards his measurement basis as $W_B=Z~(X)$. If there is a `double click' despite $W_B\in \{Z,X\}$, in which both detectors announce arrival of photon in a single line, he assigns a bit value 0 or 1 at random to his bit $b_B$, but otherwise he determines $b_B \in \{0,1\}$ depending on which detector clicks. 
If $y_B = 1$ but there is a `cross click' corresponding to multiple detections across two lines, he regards the round as $W_B = \bot$ and does not assign any bit. 
\\
(3) {\it Repetition}: Alice and Bob repeat (1) and (2) $m_{\rm rep}$ times. 
\\
(4) {\it Public communication}: For all $m_{\rm rep}$ rounds, Alice and Bob publicly announce $y_B, W_A, W_B$ and $\mu_A$. 
They also disclose $b_A$ for a round with $W_A=X$ and $b_B$ for a round with $W_B=X$. 
\\
(5) {\it Basis Reconciliation}:
From rounds with $y_B = 1$, $W_A = Z$ and $W_B = Z$, Alice and Bob randomly extract $m_{\rm sam}$ sample rounds and disclose their bits $b_A$ and $b_B$  to calculate bit error rate $e_Z$.  
Let $m_{\rm gen} := m_{\rm rep} - m_{\rm sam}$ be the number of all rounds removing rounds for the sampling, called non-sampling rounds.
Hereafter we use the notation $m(\Omega = \omega)$ as the number of non-sampling rounds where a condition $\Omega = \omega$ is satisfied. 
Alice and Bob obtain a bit sequence with length 
$m_{W, \mu'} \coloneqq m(y_B=1 \land W_A = W_B = W \land \mu_A=\mu')$ for $W \in \{Z,X\}$ and $\mu' \in \{ \mu, \nu, 0\}$. 
For $Z$ basis, they generate a sifted key with length $m_Z \coloneqq m_{Z, \mu} + m_{Z, \nu} + m_{Z, 0}$. 
For $X$ basis, they obtain the number of bit errors, 
$m_{X, \mu'}^{\rm error}  \coloneqq m(y_B=1 \land W_A = W_B = X \land \mu_A=\mu' \land b_A \neq b_B)$, for each of $\mu' \in \{ \mu, \nu, 0\}$. 
They obtain the number of cross clicks, $m_{\rm cross, \mu'}\coloneqq m(y_B=1 \land W_B = \bot  \land \mu_A = \mu')$, for each of 
$\mu' \in \{ \mu, \nu, 0\}$. 
They also obtain the number $m_{\mu'} \coloneqq m(\mu_A = \mu')$ for $\mu' \in \{ \mu, \nu, 0\}$. 
\\
(6) {\it Error correction}: Alice generates a syndrome of length $m_{\rm gen}f_{\rm EC} $ depending on the bit error rate $e_Z$ from her sifted key. 
She encrypts it by consuming a pre-shared key with Bob and sends it to him \cite{2006Koashi}.   
Bob corrects his key based on the decrypted syndrome. 
\\
(7) {\it Privacy amplification}: Alice and Bob conduct privacy amplification by shortening their keys by $m_{\rm gen} f_{\rm PA} $ bits to obtain a final key.
\\

Let us define the following ratios by using the numbers appearing in the above protocol: 
\begin{align}
Q_Z \coloneqq \frac{m_Z}{m_{\rm gen}},~Q_{Z|\mu_A}\coloneqq\frac{m_{Z, \mu_A}}{m_{\mu_A}},~
E_{X|\mu_A} \coloneqq\frac{m_{X,\mu_A}^{\rm error}}{m_{\mu_A}}, 
~Q_{\rm cross|\mu_A}\coloneqq \frac{m_{\rm cross, \mu_A}}{m_{\mu_A}}.
\label{observed}
\end{align}
In this paper, we consider the asymptotic secure key rate in the limit of $m_{\rm rep}\to \infty$. 
\noindent
In this limit, the asymptotic value of $f_{\rm PA}$ is expressed as a function of $\{ Q_{Z|\mu_A}$, $E_{X|\mu_A} $, $Q_{{\rm cross}|\mu_A}$  $\}_{\mu_A \in\{\mu, \nu, 0\}}$. 
The key rate $R$ per non-sampling round is given by
\begin{equation}
R= Q_{Z}-f_{\rm PA}(\{ Q_{Z|\mu_A}, E_{X|\mu_A} , Q_{{\rm cross}|\mu_A}  \}_{\mu_A \in\{\mu, \nu, 0\}}) -f_{\rm EC}.
\label{ki-re-}
\end{equation}

\section{Security proof} \label{Proof}
The difficulty of security proof for the passive-biased BB84 protocol is stemming from the fact that a probability of Bob's basis choice depends on the photon number ($n_B$) contained in an incoming signal.  
More details of theoretical difference between the active and the passive basis choices are shown in Appendix \ref{Difference}. 
Our approach to overcome the difficulty is to focus on rounds with $n_B=1$ to fix the probability of Bob's basis choice, which enables the proof to be reduced to that of the active basis choice. As the proof for the active basis choice, we adopt the security proof with complementarity \cite{2006Koashi, 2009Koashi}. 
To realize this scenario, the rounds where Alice emits no less than a single photon and Bob receives multiple photons $(n_A\geq1 \land n_B\geq2)$ are considered to be insecure in our proof. 
To estimate the contribution of multiple-photon detection to key generation, we use the observed number of cross clicks \cite{2024Kamin}.


\begin{figure*}[t]
 \centering
 \includegraphics[keepaspectratio, scale=0.6]
      {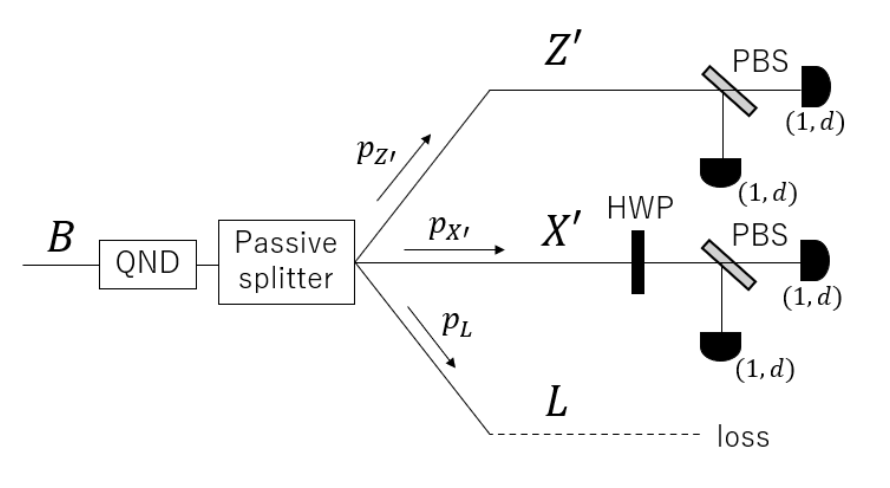}
 \caption{
 Bob's virtual setup equivalent to the actual setup shown in Fig.~\ref{setup1}. The QND measurement is conducted to obtain the photon number $n_B$. Incoming light is split into $Z'$, $X'$ and $L$ line with the ratio of $p_{Z'}\coloneqq p_Z$, $p_{X'}\coloneqq p_X r$ and $p_L \coloneqq 1-p_{Z'}-p_{X'}$, respectively, where $r = \eta_X / \eta_Z$. The $L$ line represents photon loss while $Z'$ and $X'$ lines are lossless. All threshold detectors in $Z'$ and $X'$ lines have the quantum efficiency $1$ and dark count probability $d$. 
 }
 \label{virtual_setup}
\end{figure*}

\subsection{Alternative protocol}
\begin{figure*}[t]
 \centering
 \includegraphics[keepaspectratio, scale=0.5]
      {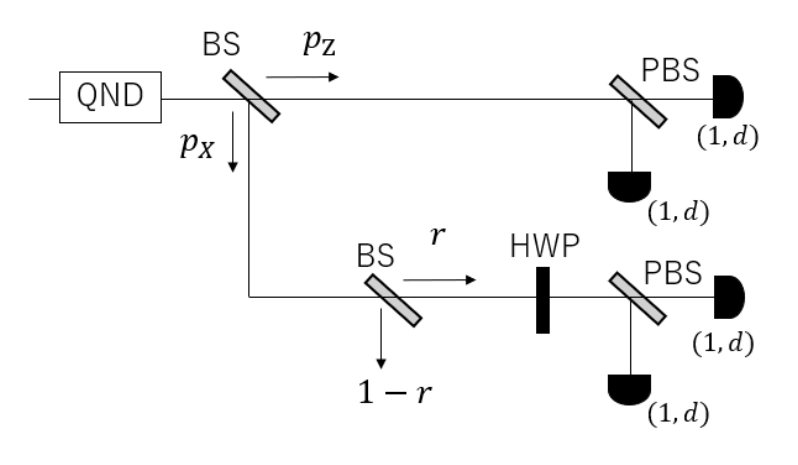}
 \caption{
Bob's virtual setups connecting Fig.~\ref{setup1} and Fig.~\ref{virtual_setup}. The QND measurement is conducted to obtain the photon number $n_B$. The common transmittance $\eta_Z$ between $Z$ and $X$ lines in Fig.~\ref{setup1} is absorbed in the quantum channel. The transmittance in $Z$ line is 1 and that in $X$ line is $r = \eta_X/\eta_Z$ modeled as a BS in the figure. All threshold detectors have the quantum efficiency $1$ and the dark count probability $d$.
 }
 \label{setup2}
\end{figure*}

Our starting point  is to introduce an alternative protocol which is equivalent to the actual protocol from the viewpoint of an eavesdropper called Eve. 
The alternative protocol is described by replacing the steps (1) and (2) in the actual protocol to the following: 

\noindent
(1') {\it State Preparation}: Alice first chooses a photon number $n_A$ according to Poisson distribution $P_{\mu_A}^{\rm Poisson}(n_A)$ and chooses her basis $W_A \in \{Z, X\}$ with probability $p_{W_A}$. She prepares $\ket{\Phi_Z^{(n_A)}}_{AA'}$, followed by the projective measurement $\{ \ket{0}\bra{0}_A , \ket{1}\bra{1}_A\}$ for $Z$ basis on system $A$ to obtain the outcome $b_A$, or prepares $\ket{\Phi_X^{(n_A)}}_{AA'}$, followed by the projective measurement $\{ \ket{+}\bra{+}_A , \ket{-}\bra{-}_A\}$ for $X$ basis, where 
\begin{equation}
\begin{split}
&\ket{\Phi_Z^{(n_A)}}_{AA'} \coloneqq (\ket{0}_A \ket{H^{(n_A)}}_{A'} + \ket{1}_A \ket{V^{(n_A)}}_{A'}) / \sqrt{2}, \\
&\ket{\Phi_X^{(n_A)}}_{AA'} \coloneqq (\ket{+}_A \ket{D^{(n_A)}}_{A'} + \ket{-}_A \ket{\bar{D}^{(n_A)}}_{A'}) / \sqrt{2},
\label{replacement}
\end{split}
\end{equation}
and \{$\ket{0}_A$, $\ket{1}_A$\} is a computational basis with $\ket{\pm}_A \coloneqq (\ket{0}_A \pm \ket{1}_A)/\sqrt{2}$.  
Alice sends system $A'$ to Bob. 
Note that for $n_A=1$, 
\be
\ket{\Phi_Z^{(1)}}_{AA'}  = \ket{\Phi_X^{(1)}}_{AA'} 
\label{single-equivalence}
\ee
holds, which implies that she prepares the same state independently of her basis choice, and thus her measurement on system $A$ can be postponed until the signal arrives at Bob's site. 
For $n_A=0$, since we have $\ket{P^{(0)}}_{A'} = \ket{\rm vac}_{A'}$ for any $P \in \{H,V,D,\bar{D}\}$, 
\be
\ket{\Phi_Z^{(0)}}_{AA'} =  \ket{+}_A \ket{\rm vac}_{A'}
\label{vacuumphase}
\ee
holds. 
\\
\noindent
(2') {\it Measurement}: As shown in Fig.~\ref{virtual_setup}, Bob conducts a polarization-independent QND measurement to obtain the photon number $n_B$. He then passively splits incoming light into three passes $Z'$ line, $X'$ line and $L$ line with the splitting ratio of $p_{Z'} \coloneqq p_Z$, $p_{X'} \coloneqq p_X r$ and $p_L \coloneqq 1 - p_{Z'}  - p_{X'}$,  respectively, where $Z'$ and $X'$ lines are lossless and have detectors with quantum efficiency 1 and dark count probability $d$ per pulse while all photons in $L$ line are lost.\\
\\
We explain these replacements in the following. 
From the viewpoint of Eve, the steps (1) and (1') are equivalent because a probability that a signal with $n_A$ photons is emitted from Alice is $P_{\mu_A}^{\rm Poisson}(n_A)$ 
and the probabilities that its polarization is $H, V, D$ and $\bar{D}$ are $p_Z /2, p_Z /2, p_X /2$ and $p_X /2$, respectively, in both steps.  
The replacement from the step (1) to (1') is called `source replacement' scheme \cite{2025Tupkary}. 
In the step (2), the outcome of the actual measurement does not change even if Bob first conducts the QND measurement because the POVM elements of the actual measurement are block-diagonalized in terms of the photon number $n_B$ \cite{2008Lutkenhaus}. 
The common transmittance $\eta_Z$ between $Z$ line and $X$ line in Fig.~\ref{setup1} can be considered to be absorbed by Eve's quantum channel while the additional transmittance $r$ (defined in Eq. (\ref{rrr})) for $X$ line is left. 
The equivalent setup to Fig.~\ref{setup1} is thus shown in Fig.~\ref{setup2}. 
We often see a scenario where the dark counts are regarded as noise in Eve's quantum channel, but it holds only when dark counts do not influence the probability of basis choice. The protocol with active basis choice and the protocol with passive {\it balanced} basis choice satisfy this condition while the protocol with passive biased basis choice, which is the target of this paper, does not satisfy the condition. Thus, dark count probability $d$ is left in Fig.~\ref{setup2} while quantum efficiency of detectors can be regarded as 1. 
Next, we divide $X$ line into two lines. As shown in Fig.~\ref{virtual_setup}, we introduce $X'$ line corresponding to the path along which light is transmitted by the beam splitter with a transmittance $r$ to reach detectors in Fig.~\ref{setup2}. We introduce $L$ line corresponding to the path along which light is reflected by the beam splitter in Fig.~\ref{setup2} and also introduce $Z'$ line corresponding to $Z$ line in Fig.~\ref{setup2}. 
Consequently, in Fig.~\ref{virtual_setup}, incoming light is split into three passes, $Z'$ line, $X'$ line and $L$ line with the splitting ratio of $p_{Z'} = p_Z$, $p_{X'} = p_X r$, and $p_L = 1 - p_{Z'}  - p_{X'}$, respectively. There is no loss in $Z'$ line and $X'$ line any more.
Thus we obtain the step (2'). 

In the alternative protocol, photon numbers $n_A$ and $n_B$ are regarded as observed numbers, 
which enables us to define the following ratios for $W \in \{Z,X\}$: 
\begin{equation}
\begin{split}
Q^{(n'_A, n'_B)} &\coloneqq m(y_B=1 \land n_A=n'_A \land n_B=n'_B) / m_{\rm gen}, \\
Q_{W}^{(n'_A, n'_B)} &\coloneqq m(y_B=1 \land W_A=W_B=W \land n_A=n'_A \land n_B=n'_B) / m_{\rm gen}, \\
E_{W}^{(n'_A, n'_B)} &\coloneqq m(y_B=1 \land W_A=W_B=W \land n_A=n'_A \land n_B=n'_B \land b_A \neq b_B) / m_{\rm gen}, \\
Q_{\rm cross}^{(n'_A, n'_B)} &\coloneqq m(y_B=1 \land W_B=\bot \land n_A=n'_A \land n_B=n'_B) / m_{\rm gen}.
\label{alternativeQE}
\end{split}
\end{equation}
In this paper, the security proof with complementarity \cite{2006Koashi, 2009Koashi} is applied where the `phase error' rate is bounded by observed values in the actual protocol. 
The phase error is defined as a bit error when Alice virtually measures system $A$ of $\ket{\Phi_Z^{(n_A)}}_{AA'}$ in the $X$ basis instead of $Z$ basis to obtain a bit value $\tilde{b}_A$ and Bob outputs a bit value $\tilde{b}_B$ as a guess for Alice's measurement outcome $\tilde{b}_A$. 
By defining 
\begin{align}
E_{\rm ph}^{(n'_A, n'_B)} \coloneqq m(y_B=1 \land W_A=W_B=Z \land n_A=n'_A \land n_B=n'_B \land \tilde{b}_A \neq \tilde{b}_B) / m_{\rm gen}, 
\label{phaseerrorratio}
\end{align}
the phase error rate is represented as  $E_{\rm ph}^{(n_A, n_B)} / Q_Z^{(n_A, n_B)}$. 
Since $X$ and $Z$ bases are complementary, the more Bob can estimate Alice's $X$-basis measurement outcomes, i.e., less phase error rate, the less Alice's $Z$-basis measurement outcomes (a secret key) are leaked to Eve. That is, smaller phase error rate implies a smaller amount of privacy amplification.  

Now consider the dependency of the phase error rate on photon numbers $n_A$ and $n_B$. 
For $n_A \geq 2$, the phase error rate is regarded as 1/2. This is physically achieved when Eve extracts one photon from two photons emitted from Alice, for example, which is called photon number splitting attack \cite{2000Brassard}. 
For $n_A=1$ and $n_B \geq 2$, in this paper, we regard the phase error rate as 1/2, i.e., arrivals of multiple photons are assumed to be totally insecure. 
For $n_A=1$ and $n_B=0$, the phase error rate is 1/2 because Bob's measurement outcome is purely originated from dark counts and he cannot estimate Alice's $X$-basis measurement outcome at all. 
For $n_A=0$, from Eq. (\ref{vacuumphase}), Bob always knows that Alice's $X$-basis measurement result is $0$, and thus the phase error rate is 0. 
For $n_A=1$ and $n_B=1$, the phase error rate is non-trivial and an upper bound is given in the following section. As a consequence, the alternative protocol is secure if the ratio of privacy amplification is 
\be
f_{\rm PA}^{\rm alt} = \sum\limits_{n_A\geq2,n_B\geq0} Q_{Z}^{(n_A, n_B)} 
+ \sum\limits_{n_B\geq2} Q_{Z}^{(1, n_B)}
+ Q_{Z}^{(1, 0)} 
+ Q_{Z}^{(1, 1)} h\left(\frac{E_{\rm ph}^{(1, 1)}}{Q_{Z}^{(1, 1)}}\right), 
\label{PAformula}
\ee
where $h(x) \coloneqq -x {\rm log}_2 x - (1-x){\rm log}_2(1-x)$ represents the binary entropy. 
However, each term is not observable in the actual protocol, and thus we derive an upper bound on $f_{\rm PA}^{\rm alt}$ using observed quantities.

\subsection{Upper bounding on parameters in $f_{\rm PA}^{\rm alt}$}
In this subsection, we show that an upper bound on $f_{\rm PA}^{\rm alt}$ is given by bounds on quantities obtained by the decoy-state method. 
The bound on $f_{\rm PA}^{\rm alt}$ is derived from upper bounds on phase error rate in rounds with single-photon detection and on the rate of multiple-photon detection. 

\subsubsection{Upper bound on single-photon phase error rate} \label{section error rate}
Here we derive an upper bound on $E_{\rm ph}^{(1, 1)}$. 
The key idea is to find a protocol with Bob's active basis choice which is equivalent to the alternative protocol. 
We define $B$ as a system before Bob's passive splitter in Fig.~\ref{virtual_setup}, and also define $Z', X'$ and $L$ as systems of $Z'$ line, $X'$ line and $L$ line, respectively. 
Let  $\hat{b}^{\dagger}_H$ and $\hat{b}^{\dagger}_V$ be creation operators of horizontal and vertical polarizations, respectively, and $\hat{b}^{\dagger}_D \coloneqq (\hat{b}^{\dagger}_H + \hat{b}^{\dagger}_V)/\sqrt{2}$ and $\hat{b}^{\dagger}_{\bar{D}} \coloneqq (\hat{b}^{\dagger}_H - \hat{b}^{\dagger}_V)/\sqrt{2}$ be operators of diagonal polarizations in system $B$. 
Similarly,  define ($\hat{z}^{\dagger}_H$, $\hat{z}^{\dagger}_V$),  ($\hat{x}^{\dagger}_H$, $\hat{x}^{\dagger}_V$) and ($\hat{l}^{\dagger}_H$, $\hat{l}^{\dagger}_V$) as sets of 
creation operators of horizontal and vertical polarizations in systems $Z', X'$ and $L$, respectively. 
We also define $\hat{x}^{\dagger}_D := (\hat{x}^{\dagger}_H + \hat{x}^{\dagger}_V)/\sqrt{2}$ and $\hat{x}^{\dagger}_{\bar{D}} := (\hat{x}^{\dagger}_H - \hat{x}^{\dagger}_V)/\sqrt{2}$. With a notation 
\begin{align}
\hat{N}_{\xi}^{(1)} &\coloneqq \hat{\xi}^{\dagger} \ket{\rm vac} \bra{\rm vac}_{Z'X'L} \hat{\xi},
\end{align}
for $\xi \in \{z_H, z_V, x_D, x_{\bar{D}}, l_H, l_V\}$, the POVM elements in system $Z'X'L$ under the condition that the outcome of QND measurement is $n_B =1$ 
are described as follows: 
\begin{equation}
\begin{split}
&\hat{F}_{Z0}^{(1)} = \left( \hat{N}_{z_H}^{(1)} + d (\hat{N}_{l_H}^{(1)} + \hat{N}_{l_V}^{(1)}) \right) (1-d)^3,
~~\hat{F}_{Z1}^{(1)} = \left( \hat{N}_{z_V}^{(1)} + d (\hat{N}_{l_H}^{(1)} + \hat{N}_{l_V}^{(1)}) \right) (1-d)^3,  \\
&\hat{F}_{Z, \rm double}^{(1)} = \Big( d ( \hat{N}_{z_H}^{(1)} +  \hat{N}_{z_V}^{(1)}) + d^2 (\hat{N}_{l_H}^{(1)} + \hat{N}_{l_V}^{(1)}) \Big) (1-d)^2, \\
&\hat{F}_{X0}^{(1)} = \left( \hat{N}_{x_D}^{(1)} + d (\hat{N}_{l_H}^{(1)} + \hat{N}_{l_V}^{(1)}) \right) (1-d)^3,
~~\hat{F}_{X1}^{(1)} = \left( \hat{N}_{x_{\bar{D}}}^{(1)} + d (\hat{N}_{l_H}^{(1)} + \hat{N}_{l_V}^{(1)}) \right) (1-d)^3,  \\
&\hat{F}_{X, \rm double}^{(1)} = \Big( d ( \hat{N}_{x_D}^{(1)} +  \hat{N}_{x_{\bar{D}}}^{(1)}) + d^2 (\hat{N}_{l_H}^{(1)} + \hat{N}_{l_V}^{(1)}) \Big) (1-d)^2, \\
&\hat{F}_{\rm no-click}^{(1)} =  (1-d)^4 (\hat{N}_{l_H}^{(1)} + \hat{N}_{l_V}^{(1)}),  \\
&\hat{F}_{\rm cross}^{(1)} = (1-(1-d)^2) \Big( \hat{N}_{z_H}^{(1)} +  \hat{N}_{z_V}^{(1)} +  \hat{N}_{x_D}^{(1)} +  \hat{N}_{x_{\bar{D}}}^{(1)}  \Big) + (1-(1-d)^2)^2 (\hat{N}_{l_H}^{(1)} + \hat{N}_{l_V}^{(1)}),  
\label{POVMoriginals}
\end{split}
\end{equation}
with $\hat{F}_{W'b}^{(1)}$ corresponding to the outcome $b\in\{0, 1\}$ in $W'\in\{Z', X'\}$ line, 
$\hat{F}_{W', \rm double}^{(1)}$ corresponding to the double-click events in $W'\in\{Z', X'\}$ line, $\hat{F}_{\rm no-click}^{(1)}$ corresponding to the no-detection event and $\hat{F}_{\rm cross}^{(1)}$ corresponding to cross-click events. 
Note that the sum of the above POVM elements equals the projection onto the single-photon subspace of system $Z'X'L$
\begin{align}
\hat{\mathbbm{1}}_{Z'X'L}^{(1)} \coloneqq \hat{N}_{z_H}^{(1)} +  \hat{N}_{z_V}^{(1)} +  \hat{N}_{x_D}^{(1)} +  \hat{N}_{x_{\bar{D}}}^{(1)} + \hat{N}_{l_H}^{(1)} + \hat{N}_{l_V}^{(1)}.
\end{align} 

Now we rewrite those POVM elements as operators on system $B$ instead of system $Z'X'L$. 
An ancillary system $EE'$ is introduced so that the dimension of $BEE'$ equals that of $Z'X'L$. 
Let $\hat{U}$ be a unitary operator from $BEE'$ to $Z'X'L$ representing evolution of the passive splitter in Fig.~\ref{virtual_setup}, whose action is  
\begin{align}
\hat{U} \hat{b}^{\dagger}_H \hat{U}^{\dagger} = \sqrt{p_{Z'}} \hat{z}^{\dagger}_H + \sqrt{p_{X'}} \hat{x}^{\dagger}_H + \sqrt{p_L} \hat{l}^{\dagger}_H, 
~~\hat{U} \hat{b}^{\dagger}_V \hat{U}^{\dagger} = \sqrt{p_{Z'}} \hat{z}^{\dagger}_V+ \sqrt{p_{X'}} \hat{x}^{\dagger}_V + \sqrt{p_L} \hat{l}^{\dagger}_V. \label{HVevolution}
\end{align}
Here we define a projection operator $\hat{M}_{B,H}^{(1)}$ from the single-photon subspace on system $B$ onto $\hat{z}^{\dagger}_H \ket{\rm vac}_{Z'X'L}$, which is described as $\hat{M}_{B,H}^{(1)} =  {}_{Z'X'L}\bra{\rm vac} \hat{z}_H \hat{U} \ket{\rm vac}_{EE'}$. 
Let us also define
\begin{align}
\ket{H}_B \coloneqq \hat{b}^{\dagger}_H \ket{\rm vac}_B, ~~\ket{V}_B \coloneqq  \hat{b}^{\dagger}_V \ket{\rm vac}_B, 
~~\ket{D}_B  \coloneqq \hat{b}^{\dagger}_D \ket{\rm vac}_B,~~\ket{\bar{D}}_B \coloneqq \hat{b}^{\dagger}_{\bar{D}} \ket{\rm vac}_B.
\end{align}
By denoting $\hat{M}_{B,H}^{(1)}  \ket{H}_B \eqqcolon \gamma_{HH}$ and $\hat{M}_{B,H}^{(1)} \ket{V}_B \eqqcolon \gamma_{HV}$, 
\be
\hat{M}_{B,H}^{(1)} = \hat{M}_{B,H}^{(1)} \hat{\mathbbm{1}}_B^{(1)}  = {}_B \bra{H} \gamma_{HH} +  {}_B \bra{V} \gamma_{HV}
\label{MBH}
\ee
holds, where we defined $\hat{\mathbbm{1}}_B^{(1)} \coloneqq \ket{H} \bra{H}_B + \ket{V} \bra{V}_B$. Now we calculate $\gamma_{HH}$ and $\gamma_{HV}$ as follows: 
\begin{equation}
\begin{split}
\gamma_{HH} &= {}_{Z'X'L}\bra{\rm vac} \hat{z}_H \hat{U} \hat{b}^{\dagger}_H \ket{\rm vac}_{BEE'} 
~~= {}_{Z'X'L}\bra{\rm vac} \hat{z}_H \hat{U} \hat{b}^{\dagger}_H \hat{U}^{\dagger}  \ket{\rm vac}_{Z'X'L} \\
&= {}_{Z'X'L}\bra{\rm vac} \hat{z}_H(\sqrt{p_{Z'}} \hat{z}^{\dagger}_H + \sqrt{p_{X'}} \hat{x}^{\dagger}_H + \sqrt{p_L} \hat{l}^{\dagger}_H)\ket{\rm vac}_{Z'X'L}  \\
&= \sqrt{p_{Z'}}, \\
\gamma_{HV}
&= {}_{Z'X'L}\bra{\rm vac} \hat{z}_H(\sqrt{p_{Z'}} \hat{z}^{\dagger}_V + \sqrt{p_{X'}} \hat{x}^{\dagger}_V + \sqrt{p_L} \hat{l}^{\dagger}_V)\ket{\rm vac}_{Z'X'L}  \\
&= 0. 
\end{split}
\end{equation}
Therefore, from Eq. (\ref{MBH}), we obtain $\hat{M}_{B,H}^{(1)} =   {}_B\bra{H} \sqrt{p_{Z'}}$, and thus the POVM element on system $B$ is 
${\hat{M}_{B,H}^{(1) \dagger}} \hat{M}_{B,H}^{(1)} = p_{Z'}  \ket{H} \bra{H}_B$. 
Similarly, we have the following relations: 
\begin{equation}
\begin{split}
&\hat{N}_{z_H}^{(1)} \to p_{Z'} \ket{H}\bra{H}_B, ~~
\hat{N}_{z_V}^{(1)}  \to p_{Z'} \ket{V}\bra{V}_B, ~~
\hat{N}_{x_D}^{(1)}  \to p_{X'} \ket{D}\bra{D}_B,  \\
&\hat{N}_{x_{\bar{D}}}^{(1)}   \to p_{X'} \ket{\bar{D}}\bra{\bar{D}}_B, ~~
\hat{N}_{l_H}^{(1)} \to p_L \ket{H}\bra{H}_B, ~~
\hat{N}_{l_V}^{(1)} \to p_L \ket{V}\bra{V}_B.
\end{split}
\end{equation}
From these results, POVM elements in Eq. (\ref{POVMoriginals}) are represented with operators on system $B$ as follows:
\begin{equation}
\begin{split}
&\hat{F'}_{Z0}^{(1)} = \left( p_{Z'}  \ket{H} \bra{H}_B + p_L d \hat{\mathbbm{1}}_B^{(1)} \right) (1-d)^3, ~~
\hat{F'}_{Z1}^{(1)} = \left( p_{Z'}  \ket{V} \bra{V}_B + p_L d \hat{\mathbbm{1}}_B^{(1)} \right) (1-d)^3,  \\
&\hat{F'}_{Z, \rm double}^{(1)} = (p_{Z'} d + p_L d^2) (1-d)^2 \hat{\mathbbm{1}}_B^{(1)},  \\
&\hat{F'}_{X0}^{(1)} = \left( p_{X'}  \ket{D} \bra{D}_B + p_L d \hat{\mathbbm{1}}_B^{(1)} \right) (1-d)^3, ~~
\hat{F'}_{X1}^{(1)} = \left( p_{X'}  \ket{\bar{D}} \bra{\bar{D}}_B + p_L d \hat{\mathbbm{1}}_B^{(1)} \right) (1-d)^3,  \\
&\hat{F'}_{X, \rm double}^{(1)} = (p_{X'} d + p_L d^2) (1-d)^2 \hat{\mathbbm{1}}_B^{(1)},  \\
&\hat{F'}_{\rm no-click}^{(1)} = p_L (1-d)^4 \hat{\mathbbm{1}}_B^{(1)},  \\
&\hat{F'}_{\rm cross}^{(1)} = \left( 1-\Big(1+2p_Ld - p_L d^2\Big)(1-d)^2 \right) \hat{\mathbbm{1}}_B^{(1)}.
\end{split}
\end{equation}
Note that the sum of those elements equals $ \hat{\mathbbm{1}}_B^{(1)}$. 
Including classical post-processing, the outcomes are limited to `0', `1' and `failure', where `failure' corresponds to the no-detection or cross-click round. 
Since a random bit is assigned to a double-click round, POVM elements including the post-processing are as follows:
\begin{equation}
\begin{split}
&\hat{G}_{Z0}^{(1)} = \hat{F'}_{Z0}^{(1)} + \frac{1}{2} \hat{F'}_{Z, \rm double}^{(1)}, ~~
\hat{G}_{Z1}^{(1)} = \hat{F'}_{Z1}^{(1)} + \frac{1}{2} \hat{F'}_{Z, \rm double}^{(1)}, ~~
\hat{G}_{X0}^{(1)} = \hat{F'}_{X0}^{(1)} + \frac{1}{2} \hat{F'}_{X, \rm double}^{(1)}, \\
&\hat{G}_{X1}^{(1)} = \hat{F'}_{X1}^{(1)} + \frac{1}{2} \hat{F'}_{X, \rm double}^{(1)}, ~~
\hat{G}_{\rm fail}^{(1)} =  \hat{F'}_{\rm no-click}^{(1)} + \hat{F'}_{\rm cross}^{(1)} . 
\label{POVMs}
\end{split}
\end{equation}
Furthermore, since 
\begin{align}
\hat{G}_{Z0}^{(1)} + \hat{G}_{Z1}^{(1)} = (p_{Z'} + 2 p_L d - p_L d^2)(1-d)^2 \hat{\mathbbm{1}}_B^{(1)}, ~~
\hat{G}_{X0}^{(1)} + \hat{G}_{X1}^{(1)} = (p_{X'} + 2 p_L d - p_L d^2) (1-d)^2 \hat{\mathbbm{1}}_B^{(1)}
\end{align}
hold, the probabilities that $Z$-basis and $X$-basis measurements are conducted at Bob's site in a round with $n_B=1$ are 
$s_Z \coloneqq (p_{Z'} + 2 p_L d - p_L d^2)(1-d)^2$ and $s_X \coloneqq (p_{X'} + 2 p_L d - p_L d^2)(1-d)^2$, respectively, independently of the state given by Eve. 
Therefore, the measurement with POVM elements in Eq. (\ref{POVMs}) are realized by the protocol where Bob filters an incoming signal with the success probability $s_Z + s_X$ and {\it actively} choose  
$Z/X$ basis with probability 
\be
\tilde{p}_Z \coloneqq \frac{s_Z}{s_Z+s_X},~~\tilde{p}_X \coloneqq \frac{s_X}{s_Z+s_X},
\ee
respectively, followed by the measurement $\{ \hat{G}_{Z0}^{(1)}/ s_Z, \hat{G}_{Z1}^{(1)}/ s_Z \}$ for $Z$ basis and $\{ \hat{G}_{X0}^{(1)}/ s_X, \hat{G}_{X1}^{(1)}/ s_X \}$ for $X$ basis to deterministically obtain the outcome `0' or `1'. Here we note that $s_Z \geq s_X$ holds from $p_{Z'} \geq p_{X'}$, which leads to 
\be
\tilde{p}_Z \geq \tilde{p}_X. \label{tildepineq}
\ee

In this active protocol, Bob is allowed to perform $X$-basis measurement after the successful filtering even if he chooses $W_B=Z$, and then Bob's guess $\tilde{b}_B$ in the definition of $E_{\rm ph}^{(1,1)}$ in Eq.~(\ref{phaseerrorratio}) is defined as the output of this $X$-basis measurement. Notice that this measurement on the filtered state is equivalent to the $X$-basis measurement performed under the choice of $W_B=X$. On the other hand, in a round with $n_A =1$, from Eq. (\ref{single-equivalence}), Alice sends system $A'$ in state $\ket{\Phi_Z^{(1)}}_{AA'}$ to Bob independently of her basis choice and anyway performs $X$-basis measurement to estimate $E_X^{(1,1)}$ for the choice of $W_A=X$ and $E_{\rm ph}^{(1,1)}$ for the choice of $W_A=Z$. Hence, by considering that Alice chooses $Z$ $(X)$ basis with probability $p_Z$ $(p_X)$ and Bob chooses $Z$ $(X)$ basis with probability $\tilde{p}_Z$ $(\tilde{p}_X)$, 
we have 
\be
E_{\rm ph}^{(1,1)} \sim \frac{p_Z\tilde{p}_Z}{p_X\tilde{p}_X} E_X^{(1,1)} \label{Eph11}
\ee
in the asymptotic limit of $m_{\rm rep}\to \infty$. Here `$\sim$' denotes asymptotic equivalence.

Next we derive an upper bound on $E_{\rm ph}^{(1,1)}$ with Eq. (\ref{Eph11}). 
By defining a quantity which can be estimated with the decoy-state method as 
\be
E_X^{(n_A =1)} \coloneqq \sum\limits_{n_B \geq 0} E_X^{(1,n_B)}, 
\ee
we obtain 
\begin{align}
E_X^{(1,1)} = E_X^{(n_A =1)} - E_X^{(1,0)}  - \sum\limits_{n_B \geq 2} E_X^{(1,n_B)}  \leq E_X^{(n_A =1)} - E_X^{(1,0)}. \label{ineqEX}
\end{align}
Since a bit in a round with $n_B =0$ is originated from dark counts of detectors and the dark count rates are the same between both detectors in $X'$ line, the bit error occurs with the probability 1/2, which leads to 
\begin{align}
E_X^{(1,0)} \sim \frac{1}{2} Q_X^{(1,0)} \label{halferror}
\end{align}
in the asymptotic limit. 
On the other hand, 
\be
Q_X^{(1, 0)} \sim \frac{p_X}{p_Z} Q_Z^{(1, 0)} 
\label{QX10}
\ee
holds in the asymptotic limit because Alice can be deemed to independently choose her basis at the end for the choice of $n_A = 1$ from Eq. (\ref{single-equivalence}), while all detectors in $Z'$ line and in $X'$ line have the same dark count rate.
From Eqs. (\ref{Eph11}), (\ref{ineqEX}), (\ref{halferror}) and (\ref{QX10}), we have
\begin{align}
E_{\rm ph}^{(1,1)} \lesssim \frac{p_Z\tilde{p}_Z}{p_X\tilde{p}_X} \left(E_X^{(n_A =1)} - \frac{1}{2} \frac{p_X}{p_Z} Q_Z^{(1, 0)}\right)
 = \frac{p_Z\tilde{p}_Z}{p_X\tilde{p}_X} E_X^{(n_A =1)} - \frac{1}{2} \frac{\tilde{p}_Z}{\tilde{p}_X} Q_Z^{(1, 0)},
\label{Ebound}
\end{align} 
where `$\lesssim$' is used for an inequality which holds in the asymptotic limit. 

Here we derive an upper bound on the quantity $f_{\rm PA}^{\rm alt}$ of privacy amplification in Eq. (\ref{PAformula}) by using Eq. (\ref{Ebound}). 
Let us define 
\be
Q_Z^{(n_A=1)}\coloneqq \sum\limits_{n_B \geq 0} Q_Z^{(1, n_B)}, 
\ee
which can be estimated with the decoy-state method and define $q_Z \coloneqq Q_Z^{(n_A=1)} - \sum_{n_B\geq 2} Q_Z^{(1, n_B)}~(\geq 0)$. 
Since $Q_{Z}^{(1, 1)} = q_Z - Q_{Z}^{(1, 0)}$ holds, we have  
\begin{equation}
\begin{split}
\frac{Q_{Z}^{(1, 0)}  + Q_{Z}^{(1, 1)} h(E_{\rm ph}^{(1, 1)}/Q_{Z}^{(1, 1)})}{q_Z} 
=&\frac{Q_{Z}^{(1, 0)}}{q_Z} h\left( \frac{1}{2}\right) + \frac{q_Z - Q_{Z}^{(1, 0)}}{q_Z} h \left (\frac{E_{\rm ph}^{(1, 1)}} {q_Z- Q_{Z}^{(1, 0)}} \right) 
\leq  h \left(\frac{Q_{Z}^{(1, 0)}}{2q_Z} + \frac{E_{\rm ph}^{(1, 1)}}{q_Z} \right)  \\
\lesssim&  h \left(\frac{Q_{Z}^{(1, 0)}}{2q_Z} + \frac{p_Z\tilde{p}_Z}{q_Z p_X\tilde{p}_X} E_X^{(n_A =1)} - \frac{Q_Z^{(1, 0)}}{2q_Z} \frac{\tilde{p}_Z}{\tilde{p}_X} \right) 
=  h \left(\frac{Q_{Z}^{(1, 0)}}{2q_Z} \Big(1 - \frac{\tilde{p}_Z}{\tilde{p}_X}\Big) + \frac{p_Z\tilde{p}_Z}{q_Z p_X\tilde{p}_X} E_X^{(n_A =1)} \right) \\
\leq & h \left( \frac{p_Z\tilde{p}_Z}{q_Z p_X\tilde{p}_X} E_X^{(n_A =1)} \right), \label{longineq}
\end{split}
\end{equation} 
where the first inequality uses concavity of the binary entropy, the second inequality uses Eq. (\ref{Ebound}) and the third inequality uses $\tilde{p}_Z / \tilde{p}_X \geq 1$ from Eq. (\ref{tildepineq}). In addition, the second and the third inequalities also use the property of the binary entropy $h(x)$ such that it is a monotonically increasing function with $x$ in range $0 \leq x \leq 1/2$. To use this property, the sufficient condition is 
\be
\frac{p_Z\tilde{p}_Z}{p_X\tilde{p}_X} E_X^{(n_A =1)}   \lesssim \frac{q_Z}{2} \label{insideh}
\ee
and we derive the condition for observed parameters in the actual protocol to satisfy Eq. (\ref{insideh}) in Sec. \ref{keyrate section}. 
Let us define 
\be
Q_Z^{(n_A=0)}\coloneqq \sum\limits_{n_B\geq 0} Q_Z^{(0, n_B)},
\ee
which can be estimated with the decoy-state method. 
The quantity  $f_{\rm PA}^{\rm alt}$ of privacy amplification Eq. (\ref{PAformula}) is upper bounded as follows: 
\begin{equation}
\begin{split}
f_{\rm PA}^{\rm alt} &= \sum\limits_{n_A\geq2,n_B\geq0} Q_{Z}^{(n_A, n_B)} 
+ \sum\limits_{n_B\geq2} Q_{Z}^{(1, n_B)}
+ Q_{Z}^{(1, 0)} 
+ Q_{Z}^{(1, 1)} h\left(\frac{E_{\rm ph}^{(1, 1)}}{Q_{Z}^{(1, 1)}}\right)  \\
& \lesssim \sum\limits_{n_A\geq2,n_B\geq0} Q_{Z}^{(n_A, n_B)} 
+ \sum\limits_{n_B\geq2} Q_{Z}^{(1, n_B)}
+ q_Z h \left( \frac{p_Z\tilde{p}_Z}{q_Z p_X\tilde{p}_X} E_X^{(n_A =1)} \right)   \\
& = Q_Z  - Q_{Z}^{(n_A=0)} - Q_{Z}^{(n_A=1)} + \sum\limits_{n_B\geq2} Q_Z^{(1, n_B)} + q_Z h \left( \frac{p_Z\tilde{p}_Z}{q_Z p_X\tilde{p}_X} E_X^{(n_A =1)} \right) \\
&  = Q_Z  - Q_{Z}^{(n_A=0)} - q_Z\left(1 -  h \left( \frac{p_Z\tilde{p}_Z}{q_Z p_X\tilde{p}_X} E_X^{(n_A =1)} \right)\right) \\
& = Q_Z - Q_Z^{(n_A=0)}  - \left(Q_Z^{(n_A=1)} - \sum\limits_{n_B\geq 2} Q_Z^{(1, n_B)}\right) \left(1- h \left( \frac{p_Z\tilde{p}_Z E_X^{(n_A =1)}}{p_X\tilde{p}_X (Q_Z^{(n_A=1)} - \sum\limits_{n_B\geq 2} Q_Z^{(1, n_B)})}  \right)\right).
\label{fPAZ}
\end{split}
\end{equation}
This inequality implies that an upper bound on  $\sum_{n_B\geq 2} Q_Z^{(1, n_B)}$ gives a further upper bound on $f_{\rm PA}^{\rm alt}$. 

\subsubsection{Upper bound on multiple-photon contribution} \label{upper multi}
Here we derive an upper bound on $\sum_{n_B\geq2} Q_Z^{(1, n_B)}$ in Eq. (\ref{fPAZ}) by considering cases with given $n_B(\geq 2)$. 
The key idea is to estimate the contribution of multiple photons to key generation by using the number of cross-click events which are observed uniquely in the passive protocol \cite{2024Kamin}.
Let us define the following Fock states on system $B$ and system $Z'X'L$:
 \begin{align}
 \ket{\vec{m}_B}_B &= \ket{m_H, m_V}_B \coloneqq \frac{(\hat{b}_H^{\dagger})^{m_H} (\hat{b}_V^{\dagger})^{m_V}}{\sqrt{m_H! m_V!}} \ket{\rm vac}_B, \\
 \ket{\vec{m}_{Z'X'L}}_{Z'X'L}  &= \ket{m_{ZH}, m_{ZV},  m_{XH},  m_{XV},  m_{LH},  m_{LV}}_{Z'X'L}  \nonumber \\
 &\coloneqq \frac{(\hat{z}_H^{\dagger})^{m_{ZH}} (\hat{z}_V^{\dagger})^{m_{ZV}} (\hat{x}_H^{\dagger})^{m_{XH}} (\hat{x}_V^{\dagger})^{m_{XV}} (\hat{l}_H^{\dagger})^{m_{LH}} (\hat{l}_V^{\dagger})^{m_{LV}}}{\sqrt{m_{ZH}! m_{ZV}!  m_{XH}!  m_{XV}!  m_{LH}!  m_{LV}!}}   \ket{\rm vac}_{Z'X'L}.  \label{ZXLbasis}
 \end{align}
 We denote the following four POVM elements on system $Z' X' L$: 
 an element $\hat{F}_W^{(n_B)}$ corresponding to key generation in $W \in \{Z, X\}$ basis, 
 an element $\hat{F}_{\rm no-click}^{(n_B)}$ corresponding to the no-detection event and 
 an element $\hat{F}_{\rm cross}^{(n_B)}$ corresponding to cross-click events.
By using the basis states in Eq. (\ref{ZXLbasis}) with a notation $P(\ket{\cdot}) \coloneqq \ket{\cdot}\bra{\cdot}$, the element $\hat{F}_Z^{(n_B)}$ is described as
\begin{equation}
\begin{split}
&\hat{F}_Z^{(n_B)} = 
 (1-d)^2\bigg(  \sum\limits_{m_H = 0}^{n_B}  \sum\limits_{m_{ZH}=0}^{m_H}  \sum\limits_{m_{ZV}=0}^{n_B - m_H}  
P(\ket{m_{ZH}, m_{ZV},  0, 0,  m_H-m_{ZH},  n_B-m_H-m_{ZV}}_{Z'X'L})   \\ 
& ~~~~~~~~~~~ -  \sum\limits_{m_H=0}^{n_B} (1-d)^2 P(\ket{0, 0,  0, 0,  m_{H},  n_B - m_H}_{Z'X'L}) \bigg),   \label{FZnorg} 
\end{split}
\end{equation}
where the second term in Eq. (\ref{FZnorg}) corresponds to an event where all photons go to $L$ line and no dark count occurs in $Z'$ line. 
Similarly, the other elements are described as follows:
\begin{align}
&\hat{F}_X^{(n_B)} = (1-d)^2\bigg(  \sum\limits_{m_H = 0}^{n_B}  \sum\limits_{m_{XH}=0}^{m_H}  \sum\limits_{m_{XV}=0}^{n_B - m_H} 
P(\ket{0, 0, m_{XH}, m_{XV}, m_H-m_{XH},  n_B-m_H-m_{XV}}_{Z'X'L})   \nonumber \\ 
&~~~~~~~~~~~ -  \sum\limits_{m_H=0}^{n_B} (1-d)^2 P(\ket{0, 0,  0, 0,  m_{H},  n_B - m_H}_{Z'X'L}) \bigg),   \label{FXnorg} \\
&\hat{F}_{\rm no-click}^{(n_B)} = (1-d)^4 \sum\limits_{m_H=0}^{n_B} P(\ket{0, 0,  0, 0,  m_{H},  n_B - m_H}_{Z'X'L}), \label{Flossnorg} \\
&\hat{F}_{\rm cross}^{(n_B)} = \hat{\mathbbm{1}}_{Z'X'L}^{(n_B)} - \hat{F}_Z^{(n_B)} -\hat{F}_X^{(n_B)} - \hat{F}_{\rm no-click}^{(n_B)}, \label{Fcrossnorg}
\end{align}
where 
\begin{align}
\hat{\mathbbm{1}}_{Z'X'L}^{(n_B)} \coloneqq \sum_{\vec{m}_{Z'X'L}~\in~\{\vec{m}_{Z'X'L}~|~m _{ZH}+m_{ZV}+m_{XH}+m_{XV}+m_{LH}+m_{LV}=n_B\}} 
P(\ket{\vec{m}_{Z'X'L}}_{Z'X'L}). 
\end{align}
Again we introduce an ancillary system $EE'$ so that the dimension of $BEE'$ equals that of system $Z'X'L$. Let $\hat{\mathbbm{1}}_{BEE'}^{(n_B)}$ be an identity operator in $n_B$-photon subspace on system $BEE'$. 
Using the unitary operator $\hat{U}$ indicating evolution given by the passive splitter, 
\be
\hat{\mathbbm{1}}_{BEE'}^{(n_B)} = \hat{U}^{\dagger} \hat{\mathbbm{1}}_{Z'X'L}^{(n_B)} \hat{U}
\label{identityunitary}
\ee
holds. Here we define 
\be
\hat{\mathbbm{1}}_{B}^{(n_B)} \coloneqq \sum_{\vec{m}_B~\in~\{\vec{m}_B~|~ m_H + m_V=n_B\}} 
P(\ket{\vec{m}_B}_{B}), \label{identityonnb}
\ee
and we then have 
\be
{}_{EE'} \bra{\rm vac}  \hat{U}^{\dagger} \hat{\mathbbm{1}}_{Z'X'L}^{(n_B)} \hat{U} \ket{\rm vac}_{EE'} =\hat{\mathbbm{1}}_{B}^{(n_B)}. \label{identityreduction}
\ee

Now we represent $\hat{F}_Z^{(n_B)}$ in Eq. (\ref{FZnorg}) in terms of system $B$. 
Let us define
\be
\ket{\vec{m}_{Z'L}}_{Z'X'L} \coloneqq \ket{m_{ZH}, m_{ZV},  0, 0,  m_H-m_{ZH},  n_B-m_H-m_{ZV}}_{Z'X'L}. \label{nZL}
\ee
A measurement operator  $\hat{M}_{\vec{m}_{Z'L}}: B \to \mathbbm{C}$ which corresponds to the projection onto $\ket{\vec{m}_{Z'L}}_{Z'X'L}$ is  represented as 
$\hat{M}_{\vec{m}_{Z'L}}= {}_{Z'X'L}\bra{\vec{m}_{Z'L}} \hat{U} \ket{\rm vac}_{EE'}$. 
For arbitrary 
\be
\ket{\vec{m}'_B}_B \coloneqq \ket{m'_H, m'_V}_B, 
\ee
let $\gamma_{\vec{m}'_B, \vec{m}_{Z'L}}  \coloneqq \hat{M}_{\vec{m}_{Z'L}} \ket{\vec{m}'_B}_B.$ Then we can write $\hat{M}_{\vec{m}_{Z'L}}$ as 
\begin{align}
\hat{M}_{\vec{m}_{Z'L}} =  \hat{M}_{\vec{m}_{Z'L}}  \sum\limits_{\vec{m}'_B} \ket{\vec{m}'_B}\bra{\vec{m}'_B}_B
=\sum\limits_{\vec{m}'_B}  {}_B\bra{\vec{m}'_B} \gamma_{\vec{m}'_B, \vec{m}_{Z'L}}. \label{Kraus_multi} 
\end{align}
By using Eq. (\ref{HVevolution}), we calculate $\gamma_{\vec{m}'_B, \vec{m}_{Z'L}}$ as follows;
\begin{equation}
\begin{split}
\gamma_{\vec{m}'_B, \vec{m}_{Z'L}} &= {}_{Z'X'L}\bra{\vec{m}_{Z'L}} \hat{U} \ket{\vec{m}'}_B \ket{\rm vac}_{EE'} 
= {}_{Z'X'L}\bra{\vec{m}_{Z'L}} \frac{ \hat{U} (\hat{b}_H^{\dagger})^{m'_H} (\hat{b}_V^{\dagger})^{m'_V} } { \sqrt{m'_H! m'_V!}} \ket{\rm vac}_{BEE'}  \\
&= {}_{Z'X'L}\bra{\vec{m}_{Z'L}} \frac{ (\hat{U} \hat{b}_H^{\dagger} \hat{U}^{\dagger})^{m'_H} (\hat{U} \hat{b}_V^{\dagger} \hat{U}^{\dagger})^{m'_V} } { \sqrt{m'_H! m'_V!}} \ket{\rm vac}_{Z'X'L}  \\
&= {}_{Z'X'L}\bra{\vec{m}_{Z'L}} \frac{1 }{\sqrt{m'_H!m'_V!}} \Big(\sqrt{p_{Z'}} \hat{z}_H^{\dagger} + \sqrt{p_{X'}} \hat{x}_H^{\dagger} + \sqrt{p_L} \hat{l}_H^{\dagger} \Big)^{m'_H}   \Big(\sqrt{p_{Z'}} \hat{z}_V^{\dagger} + \sqrt{p_{X'}} \hat{x}_V^{\dagger} + \sqrt{p_L} \hat{l}_V^{\dagger} \Big)^{m'_V}  \ket{\rm vac}_{Z'X'L} \\
& = {}_{Z'X'L}\bra{\vec{m}_{Z'L}} \frac{1 }{\sqrt{m'_H!m'_V!}} 
 \sum\limits_{(i,j) \in S_{m'_H}} \frac{m'_H!}{i! j! (m'_H - i-j)!} (\sqrt{p_{Z'}} \hat{z}_H^{\dagger})^i  (\sqrt{p_{X'}} \hat{x}_H^{\dagger})^j (\sqrt{p_L} \hat{l}_H^{\dagger})^{m'_H - i-j} \\
& ~~~~~~~~~~~~~~~~~~~~~~~~~~~~~~~~~~~~~~~\sum\limits_{(i',j') \in S_{m'_V}} \frac{m'_V!}{i'! j'! (m'_V - i'-j')!} (\sqrt{p_{Z'}} \hat{z}_V^{\dagger})^{i'}  (\sqrt{p_{X'}} \hat{x}_V^{\dagger})^{j'} (\sqrt{p_L} \hat{l}_V^{\dagger})^{m'_V - i'-j'}
\ket{\rm vac}_{Z'X'L}  \\
&=  {}_{Z'X'L}\bra{\vec{m}_{Z'L}} \sum\limits_{(i,j) \in S_{m'_H}} \sum\limits_{(i',j') \in S_{m'_V}} \sqrt{\frac{m'_H!}{i! j! (m'_H - i-j)!}  \frac{m'_V!}{i'! j'! (m'_V - i'-j')!}}  \\
&~~~~~~~~~~~~~~~~~~~~~~(\sqrt{p_{Z'}})^{i+i'} (\sqrt{p_{X'}})^{j+j'} (\sqrt{p_L})^{m'_H + m'_V -i-j-i'-j'}
\ket{i, i', j, j', m'_H - i-j, m'_V - i'-j'}_{Z'X'L}, \label{gammaprocess}
\end{split}
\end{equation}
where $S_{m'_H} = \{(i, j)~|~i+j \leq m'_H\}$ and $S_{m'_V} = \{(i', j')~|~i'+j' \leq m'_V\}$. 
From Eq. (\ref{nZL}), a term  in Eq. (\ref{gammaprocess}) is not zero only if $i = m_{ZH}, i' = m_{ZV}, j = 0, j'=0, m'_H = m_H$ and $ m'_V = n_B - m_H$ hold. 
If $m'_H \neq m_H$ or $m'_V \neq n_B - m_H$, then $\gamma_{\vec{m}'_B, \vec{m}_{Z'L}} = 0$ and if $m'_H = m_H$ and $m'_V = n_B - m_H$, then 
\begin{equation}
\begin{split}
\gamma_{\vec{m}'_B, \vec{m}_{Z'L}} &=
\sqrt{\frac{m_H!}{{m_{ZH}}!  (m_H - m_{ZH})!}  \frac{(n_B - m_H)!}{m_{ZV}!  (n_B - m_H - m_{ZV})!}} 
(\sqrt{p_{Z'}})^{m_{ZH}+m_{ZV}} (\sqrt{p_L})^{n_B - m_{ZH} - m_{ZV}}  \\
& = \sqrt{\binom{m_H}{m_{ZH}}\binom{n_B - m_H}{m_{ZV}}} (\sqrt{p_{Z'}})^{m_{ZH}+m_{ZV}} (\sqrt{p_L})^{n_B - m_{ZH} - m_{ZV}}.
\end{split}
\end{equation}
From Eq. (\ref{Kraus_multi}), we obtain 
\begin{align}
\hat{M}_{\vec{m}_{Z'L}}  =   \sqrt{ \binom{m_H}{m_{ZH}} \binom{n_B-m_H}{m_{ZV}}}  (\sqrt{p_{Z'}})^{m_{ZH} + m_{ZV}}  (\sqrt{p_L})^{n_B - m_{ZH} - m_{ZV}} 
{}_B\bra{m_H, n_B - m_H}. 
\end{align}
Thus, a POVM element on system $B$ corresponding to projection onto $\ket{\vec{m}_{Z'L}}_{Z'X'L}$ is 
\begin{align}
 P({}_{EE'}\bra{\rm vac}\hat{U}^{\dagger} \ket{\vec{m}_{Z'L}}_{Z'X'L})  
= \hat{M}_{\vec{m}_{Z'L}} ^{\dagger} \hat{M}_{\vec{m}_{Z'L}} 
=   \binom{m_H}{m_{ZH}} \binom{n_B-m_H}{m_{ZV}} p_{Z'}^{m_{ZH} + m_{ZV}}  p_L^{n_B - m_{ZH} - m_{ZV}}P(\ket{m_H, n_B - m_H}_B) \label{mouikutsu}. 
\end{align}
By using Eq. (\ref{mouikutsu}), the POVM element in Eq. (\ref{FZnorg}) is described in terms of system $B$ as follows.   
\begin{equation}
\begin{split}
 \hat{F'}_Z^{(n_B)} & \coloneqq
{}_{EE'} \bra{\rm vac} \hat{U}^{\dagger} \hat{F}_Z^{(n_B)} \hat{U} \ket{\rm vac}_{EE'}  \\
&= (1-d)^2\bigg(  \sum\limits_{m_H = 0}^{n_B}  \sum\limits_{m_{ZH}=0}^{m_H}  \sum\limits_{m_{ZV}=0}^{n_B-m_H}  
\binom{m_H}{m_{ZH}} \binom{n_B-m_H}{m_{ZV}} p_{Z'}^{m_{ZH} + m_{ZV}}  p_L^{n_B - m_{ZH} - m_{ZV}}P(\ket{m_H, n_B - m_H}_B)   \\ 
& ~~~~~~~ -  \sum\limits_{m_H=0}^{n_B} (1-d)^2  p_L^{n_B}P(\ket{m_H, n_B - m_H}_B) \bigg)  \\
&= (1-d)^2\bigg(  \sum\limits_{m_H = 0}^{n_B}  \sum\limits_{m_{ZH}=0}^{m_H}  \binom{m_H}{m_{ZH}}  p_{Z'}^{m_{ZH}} p_L ^{m_H - m_{ZH} } 
 \sum\limits_{m_{ZV}=0}^{n_B-m_H}   \binom{n_B-m_H}{m_{ZV}} p_{Z'}^{m_{ZV}}  p_L^{n_B - m_H - m_{ZV}}P(\ket{m_H, n_B - m_H}_B)  
 -  (1-d)^2  p_L^{n_B} \hat{\mathbbm{1}}_{B}^{(n_B)} \bigg)  \\
& =(1-d)^2\bigg(  \sum\limits_{m_H = 0}^{n_B} (p_{Z'} + p_L)^{m_H} (p_{Z'} + p_L)^{n_B - m_H} 
P(\ket{m_H, n_B - m_H}_B) -  (1-d)^2  p_L^{n_B} \hat{\mathbbm{1}}_{B}^{(n_B)} \bigg)  \\
&= (1-d)^2 \bigg((p_{Z'} + p_L)^{n_B} - (1-d)^2  p_L^{n_B}   \bigg) \hat{\mathbbm{1}}_{B}^{(n_B)}. \label{FZn}
\end{split}
\end{equation}
In similar ways, Eqs. (\ref{FXnorg}) and (\ref{Flossnorg}) are described in terms of system $B$:
\begin{align}
\hat{F'}_X^{(n_B)} &\coloneqq 
{}_{EE'} \bra{\rm vac} \hat{U}^{\dagger} \hat{F}_X^{(n_B)}  \hat{U} \ket{\rm vac}_{EE'}  
= (1-d)^2 \Big((p_{X'}+p_L)^{n_B} - (1-d)^2 p_L^{n_B}) \Big) \hat{\mathbbm{1}}_{B}^{(n_B)}, \label{FXn}\\
\hat{F'}_{\rm no-click}^{(n_B)} & \coloneqq 
{}_{EE'} \bra{\rm vac} \hat{U}^{\dagger} \hat{F}_{\rm no-click}^{(n_B)} \hat{U} \ket{\rm vac}_{EE'}  
=(1-d)^4 p_L^{n_B}  \hat{\mathbbm{1}}_{B}^{(n_B)}. \label{Flossn}
\end{align}
By using 
Eqs. (\ref{identityreduction}), (\ref{FZn})-(\ref{Flossn}), we describe Eq. (\ref{Fcrossnorg}) in terms of system $B$;
\begin{equation}
\begin{split}
\hat{F'}_{\rm cross}^{(n_B)} &\coloneqq 
{}_{EE'} \bra{\rm vac} \hat{U}^{\dagger} \hat{F}_{\rm cross}^{(n_B)} \hat{U} \ket{\rm vac}_{EE'}  
= \hat{\mathbbm{1}}_{B}^{(n_B)} - \hat{F'}_Z^{(n_B)} -\hat{F'}_X^{(n_B)} - \hat{F'}_{\rm no-click}^{(n_B)}  \\
&=\left( 1 - (1-d)^2 \Big( (p_{Z'} + p_L)^{n_B} + (p_{X'} + p_L)^{n_B} -(1-d)^2 p_L^{n_B} \Big) \right) \hat{\mathbbm{1}}_{B}^{(n_B)}.
\label{Fcrossn}
\end{split}
\end{equation}

From Eq. (\ref{FZn}), the probability $s_{Z|n_B}$ that Bob obtains a $Z$-basis key under the condition that QND-measurement outcome is $n_B$ is 
independent of the state of system $B$, which is 
\begin{align}
s_{Z|n_B} = (1-d)^2 \left ((p_{Z'} + p_L)^{n_B} - (1-d)^2 p_L^{n_B} \right) 
\label{sZn}
\end{align}
for arbitrary $n_B$. 
From Eq. (\ref{Fcrossn}), the probability $s_{{\rm cross}|n_B}$ that the cross click occurs under the condition that QND-measurement outcome is $n_B$ is also 
independent of the state of system $B$, which is
\begin{align}
s_{{\rm cross}|n_B} = 1 - (1-d)^2 \left( (p_{Z'} + p_L)^{n_B} + (p_{X'} + p_L)^{n_B} - (1-d)^2 p_L^{n_B} \right)
\end{align}
for arbitrary $n_B$. 
In a round with Alice's photon number $n_A=1$, from Eq. (\ref{single-equivalence}), she is considered to send the state $\ket{\Phi_Z^{(1)}}_{AA'} $ to Bob regardless of her basis choice, and thus the outcome $n_B$ of Bob's QND measurement is independent of Alice's basis choice. 
On the other hand, the probability of Bob's passive basis choice only depends on photon number $n_B$ as shown in Eq. (\ref{sZn}). 
These imply that Alice's basis choice and Bob's passive basis choice are independent.  
Therefore, in a round where Alice emits a single photon and Bob receives $n_B$ photons, 
the probability that a $Z$-basis sifted key is generated is 
$p_Z s_{Z|n_B} $. 
This equals $Q_Z^{(1, n_B)}/Q^{(1, n_B)}$ in the asymptotic limit, that is, $Q_Z^{(1, n_B)} / Q^{(1, n_B)} \sim p_Z s_{Z|n_B}$. Similarly, $Q_{\rm cross}^{(1, n_B)} / Q^{(1, n_B)} \sim s_{{\rm cross}|n_B}$ also holds. 
Thus, 
\begin{equation}
\begin{split}
Q_Z^{(1, n_B)} & \sim \frac{p_Z s_{Z|n_B} }{s_{{\rm cross}|n_B}} Q_{\rm cross}^{(1, n_B)} 
= \frac{p_Z \left ((p_{Z'} + p_L)^{n_B} - (1-d)^2 p_L^{n_B} \right) (1-d)^2 Q_{\rm cross}^{(1, n_B)}}
{1 - \left( (p_{Z'} + p_L)^{n_B} + (p_{X'} + p_L)^{n_B} -(1-d)^2 p_L^{n_B} \right)(1-d)^2}  \\
& = \frac{p_Z \left ((p_{Z'} + p_L)^{n_B} - (1-d)^2 p_L^{n_B} \right)  Q_{\rm cross}^{(1, n_B)}}
{\frac{1}{(1-d)^2} - \left( (p_{Z'} + p_L)^{n_B} + (p_{X'} + p_L)^{n_B} -(1-d)^2 p_L^{n_B} \right)}  \\
& \leq  \frac{p_Z \left ((p_{Z'} + p_L)^{n_B} - (1-d)^2 p_L^{n_B} \right)  Q_{\rm cross}^{(1, n_B)}}
{1 - \left( (p_{Z'} + p_L)^{n_B} + (p_{X'} + p_L)^{n_B} -(1-d)^2 p_L^{n_B} \right)}  \\
& \leq  \frac{p_Z(p_{Z'} + p_L)^{n_B}  Q_{\rm cross}^{(1, n_B)}}
{1 - \left( (p_{Z'} + p_L)^{n_B} + (p_{X'} + p_L)^{n_B}  \right)}.
\end{split}
\end{equation}
Since $p_{Z'} + p_L<1$ and $p_{X'} + p_L <1$ hold, for any $n_B \geq 2$, we have
\be
Q_Z^{(1, n_B)} \lesssim \alpha Q_{\rm cross}^{(1, n_B)}, 
\label{crossineq}
\ee
where
\be
\alpha \coloneqq \frac{p_Z(p_{Z'} + p_L)^{2}}
{1 - \left( (p_{Z'} + p_L)^{2} + (p_{X'} + p_L)^{2} \right)}. 
\ee
By defining a quantity
\be
Q_{\rm cross}^{(n_A =1)} \coloneqq \sum\limits_{n_B\geq 0} Q_{\rm cross}^{(1, n_B)}, 
\ee
which can be estimated with the decoy-state method, Eq. (\ref{crossineq}) leads to 
\begin{align}
\sum\limits_{n_B\geq 2} Q_Z^{(1, n_B)} \lesssim  \alpha Q_{\rm cross}^{(n_A =1)}. \label{one-one-bound}
\end{align}
By substituting it to Eq. (\ref{fPAZ}), we obtain
\begin{align}
f_{\rm PA}^{\rm alt} &\lesssim Q_Z - Q_Z^{(n_A=0)}  -(Q_Z^{(n_A=1)} - \alpha Q_{\rm cross}^{(n_A =1)}) \left( 1- h \left( \frac{p_Z\tilde{p}_Z E_X^{(n_A =1)}}{p_X\tilde{p}_X (Q_Z^{(n_A=1)} - \alpha Q_{\rm cross}^{(n_A =1)})}  \right) \right).
\label{lastPA}
\end{align}

\subsection{Estimation with Decoy-state method } \label{decoy section}
Here we derive bounds on parameters $E_X^{(n_A=1)}$, $Q_Z^{(n_A=0)}$, $Q_Z^{(n_A=1)}$ and $Q_{\rm cross}^{(n_A=1)}$ in Eq. (\ref{lastPA}), which are mentioned in the previous sections as values that can be evaluated with the decoy-state method. As described in the actual protocol, the decoy-state method with weak coherent states and the vacuum is used. 
In Appendix \ref{decoyproof}, we confirm that the decoy-state method can be applied to the passive protocol in a similar way to the active protocol \cite{2005Ma}.   
A difference from the active protocol is that we estimate the cross-click rate with the decoy-state method as well. 
The result is as follows:
\begin{equation}
\begin{split}
Q_Z^{(n_A=0)}&\gtrsim  \barbelow{Q}_Z^{(n_A=0)} \coloneqq  (p_\mu e^{-\mu} + p_\nu e^{-\nu} +p_0) Q_{Z|0},
 \\
Q_Z^{(n_A=1)}&\gtrsim \barbelow{Q}_Z^{(n_A=1)} \coloneqq \frac{\mu(p_\mu e^{-\mu}\mu + p_\nu e^{-\nu}\nu)}{\mu \nu - \nu^2} (Q_{Z|\nu} e^{\nu} - Q_{Z|\mu} e^{\mu}\frac{\nu^2}{\mu^2}
- Q_{Z|0}\frac{\mu^2 - \nu^2}{\mu^2}),  \\
E_X^{(n_A=1)} &\lesssim \bar{E}_X^{(n_A=1)}
\coloneqq
(p_\mu e^{-\mu}\mu + p_\nu e^{-\nu}\nu) \frac{E_{X|\nu}e^{\nu} - E_{X|0}}{\nu},  \\
Q_{\rm cross}^{(n_A=1)} &\lesssim \bar{Q}_{\rm cross}^{(n_A=1)}
\coloneqq
(p_\mu e^{-\mu}\mu + p_\nu e^{-\nu}\nu) \frac{Q_{{\rm cross}|\nu}e^{\nu} - Q_{{\rm cross}|0}}{\nu}.
\label{decoybounds}
\end{split}
\end{equation}
We can see that all parameters in the right-hand side are known in advance or observed after the implementation.

\subsection{Secure key rate} \label{keyrate section} 
Now we derive the secure key rate based on the results so far. 
By combining Eq. (\ref{lastPA}) with Eq. (\ref{decoybounds}), if we set 
\begin{align}
f_{\rm PA} &= Q_Z - \barbelow{Q}_Z^{(n_A=0)}  
 -(\barbelow{Q}_Z^{(n_A=1)} - \alpha \bar{Q}_{\rm cross}^{(n_A =1)}) \left(1- h \left( \frac{p_Z\tilde{p}_Z \bar{E}_X^{(n_A =1)}}{p_X\tilde{p}_X (\barbelow{Q}_Z^{(n_A=1)} - \alpha \bar{Q}_{\rm cross}^{(n_A =1)})}  \right) \right),
\end{align}
the alternative protocol is secure, i.e., the actual protocol is also secure. Therefore, the key generation rate per non-sampling round is given by 
\begin{align}
R = Q_Z - f_{\rm PA} - f_{\rm EC} 
=   \barbelow{Q}_Z^{(n_A=0)} + (\barbelow{Q}_Z^{(n_A=1)} - \alpha \bar{Q}_{\rm cross}^{(n_A =1)}) \left(1- h \left( \frac{p_Z\tilde{p}_Z \bar{E}_X^{(n_A =1)}}{p_X\tilde{p}_X (\barbelow{Q}_Z^{(n_A=1)} - \alpha \bar{Q}_{\rm cross}^{(n_A =1)})}  \right)\right)  - f_{\rm EC}
\label{finalkeyrate}
\end{align}
as long as $R>0$ and the following inequality are satisfied. 
\be
\frac{p_Z\tilde{p}_Z}{p_X\tilde{p}_X} \bar{E}_X^{(n_A =1)} \leq \frac{1}{2} (\barbelow{Q}_Z^{(n_A=1)}\ - \alpha \bar{Q}_{\rm cross}^{(n_A =1)} ). 
\label{suffforpre}
\ee
Note that Eq. (\ref{suffforpre}) is a sufficient condition for Eq. (\ref{insideh}) because $E_X^{(n_A =1)} \lesssim \bar{E}_X^{(n_A =1)}$ and $q_Z = Q_Z^{(n_A=1)}\ - \sum_{n_B\geq 2} Q_Z^{(1, n_B)} \gtrsim \barbelow{Q}_Z^{(n_A=1)}\ - \alpha \bar{Q}_{\rm cross}^{(n_A =1)}$ hold from Eqs. (\ref{one-one-bound}) and (\ref{decoybounds}).

\section{Numerical analysis} \label{numerical analysis}
In this section, we simulate the secure key rate with realistic parameters for experiments to compare the active and the passive protocol. We also show that the bounds in our security proof are tight enough under such a realistic situation. 

\begin{figure}[t]
 \centering
 \includegraphics[keepaspectratio, scale=0.4]
      {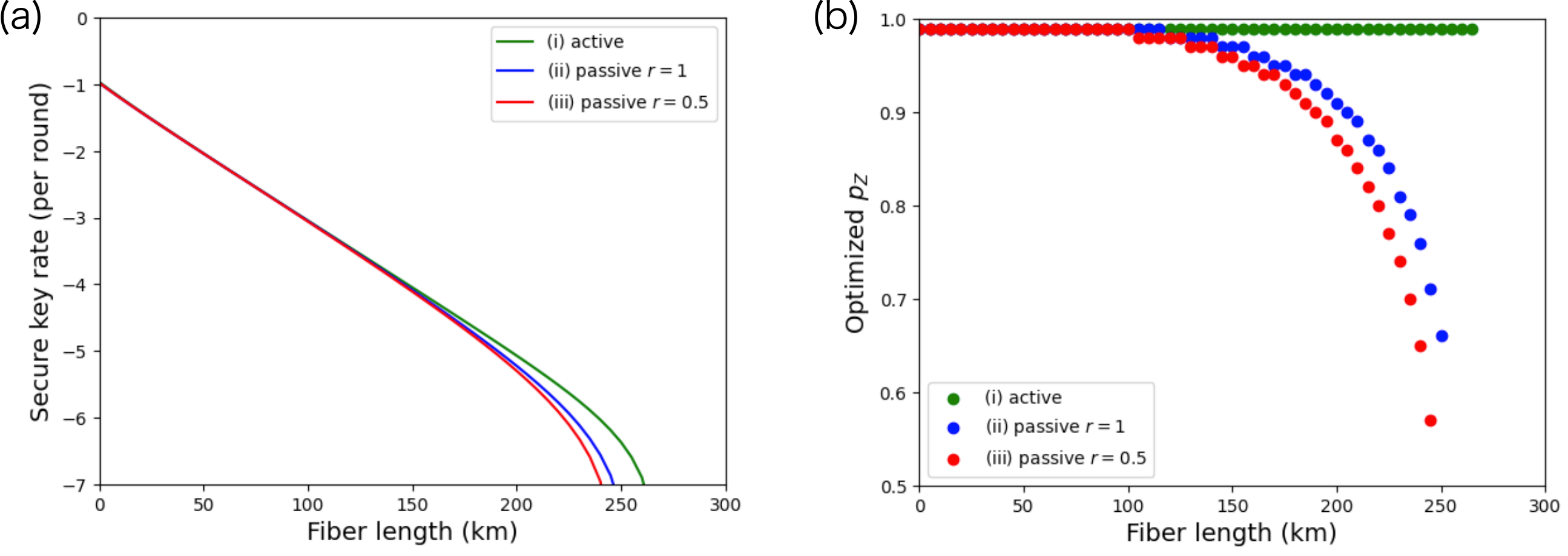}
 \caption{Secure key rate and the optimal probability of the basis choice as a function of fiber length with the quantum efficiency $\eta_{\rm det}^Z = \eta_{\rm det}^X$=0.7 and the dark count probability $d=10^{-7}$ of detectors. (a) Secure key rate $R$ per round on a logarithmic scale with base 10 as a function of fiber length (km). The top line (green) is the secure key rate of active-biased protocol for comparison. The middle line (blue) is our result $R$ for the passive-biased protocol with $r=1$ representing the symmetry of transmittance between $Z$ line and $X$ line. 
The bottom line (red) is $R$ with $r=1/2$. We assume that the intensity of a weak coherent decoy state is $\nu = 0.05$, the channel error rate is $e_{\rm ch} = 0.03$ and the error-correction coefficient is $c_{\rm EC}=1.1$. The probability $p_Z$ and the signal intensity $\mu$ are optimized at each distance. (b) Optimal $p_Z$ as a function of fiber length (km). The value $p_Z$ is optimized in the range of $[0.5, 0.99]$ in increments of $0.01$. 
 }
 \label{keyrategraph}
\end{figure}

\begin{figure}
 \centering
 \includegraphics[keepaspectratio, scale=0.4]
      {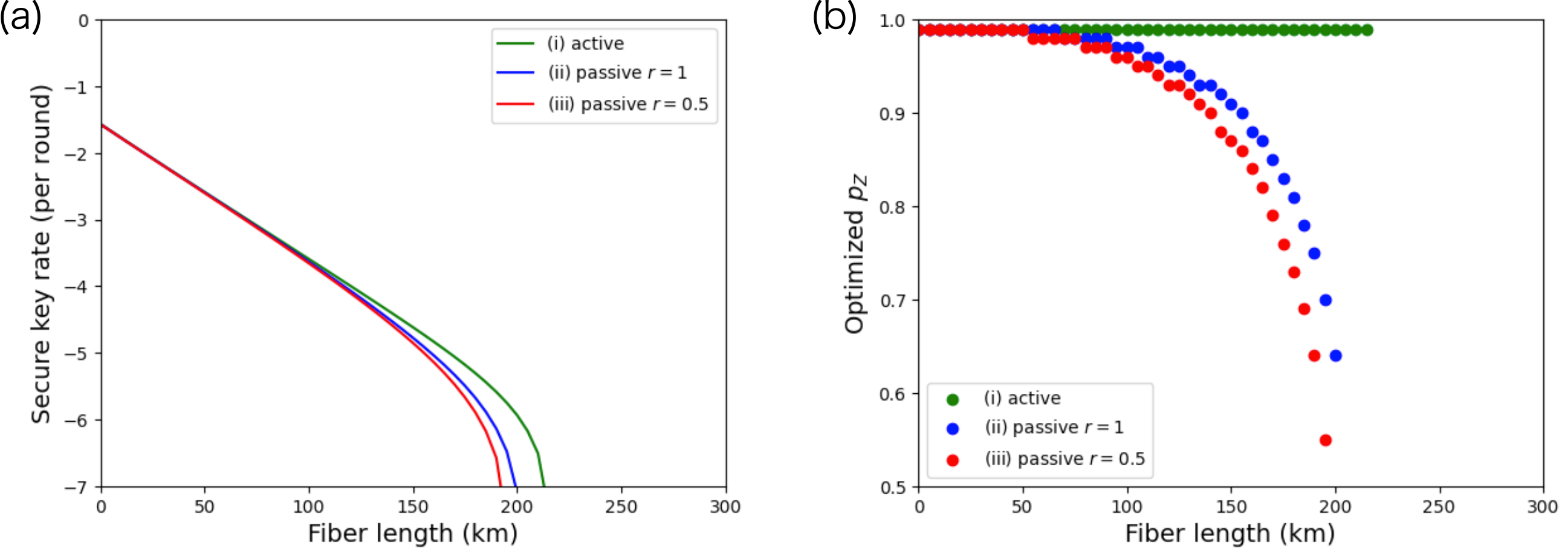}
 \caption{Secure key rate and the optimal probability of the basis choice as a function of fiber length with the quantum efficiency $\eta_{\rm def}^Z = \eta_{\rm det}^X$=0.2 and the dark count probability $d=3 \times10^{-7}$ of detectors. The other conditions are the same as those in Fig.~\ref{keyrategraph}.
 }
 \label{keyrategraph2}
\end{figure}

\subsection{Simulation of secure key rate} \label{simulation}
We show results of numerical calculation of the key rate $R$ per non-sampling round given by Eq. (\ref{finalkeyrate}). 
In the simulation shown in Fig.~\ref{keyrategraph}, we assume detectors' quantum efficiency $\eta_{\rm det}^Z = \eta_{\rm det}^X = 0.7$ and dark count probability $d = 10^{-7}$, which are achievable with commercial SSPDs \cite{2024Sanari}. 
The overall transmittances of $Z$ line and $X$ line are $\eta_Z = \eta_{\rm det}^Z$ and $\eta_X = \eta_{\rm det}^X r$, respectively. 
The channel transmittance of the optical fiber $\eta_{\rm ch}$ connecting Alice and Bob is modeled as $\eta_{\rm ch} = 10^{-l/50}$ where $l$ indicates its length (km). 
The bit error rates $e_Z$ and $e_X$ of $Z$ basis and $X$ basis, respectively, are calculated from the dark count rate in addition to the channel error rate $e_{\rm ch} = 0.03$. 
The cost for error correction $f_{\rm EC}$ is $c_{\rm EC} Q_{Z} h(e_Z)$ where the error-correction coefficient $c_{\rm EC}$ is set to be 1.1~\cite{2024Luo}. 
Regarding the decoy-state method, we assume $p_{\mu} \to 1$ because ratios using the decoy intensities $\nu$ and $0$ can be negligibly small in the asymptotic limit of $m_{\rm rep} \to \infty$. 
The decoy intensity $\nu$ is fixed to be 0.05. Optimizing signal intensity $\mu$ and the probability $p_Z$ at each distance,  
in Fig.~\ref{keyrategraph} (a), we plot numerical results  for three cases: (i) a secure key rate of the active-biased BB84 protocol for comparison (green), (ii) our result $R$ with $r = 1$ (blue) and (iii) our result $R$ with $r = 0.5$ (red). The vertical axis is on a logarithmic scale with base 10.
In Fig.~\ref{keyrategraph} (b), we plot the optimal values of  $p_Z$ as a function of the fiber length. The value $p_Z$ is optimized in the range of $[0.5, 0.99]$ in increments of $0.01$. 

Furthermore, assuming a more conservative scenario, in Fig.~\ref{keyrategraph2} (a) and Fig.~\ref{keyrategraph2} (b), we plot numerical results under the same conditions as in Fig.~\ref{keyrategraph} except with the quantum efficiency $\eta_{\rm det}^Z = \eta_{\rm det}^X = 0.2$ and the dark count probability $d = 3 \times 10^{-7}$, which are achievable with current SPAD technology \cite{2022Li}. 

As shown in both Fig.~\ref{keyrategraph} and Fig.~\ref{keyrategraph2}, there is an intrinsic gap between active case (i) and passive case (ii) which implicates the vulnerability of the passive basis choice to dark counts. 
To explain this, let us consider the signal to noise (S/N) ratio where noise is generated from dark counts of detectors. 
The secure key cannot be generated when the order of the S/N ratio is 1 because it induces a large error rate. 
For the active-biased protocol, in which the same setup is used for $Z$ basis and $X$ basis except for the wave plate to change the basis, the S/N ratio is independent of the basis choice. 
On the other hand, for the passive-biased protocol, in which a signal is divided into $Z$ line and $X$ line corresponding to the passive basis choice, the signal (and thus the S/N ratio) in $X$ basis is smaller than the active case by the factor of $p_X r$. 
Thus, in the range of long distances where the dark count rate is comparable to the signal in $X$ basis, to avoid the S/N ratio approaching to the order of 1, one must make $p_X$ larger, i.e., $p_Z$ smaller, which leads to lower key generation rate in $Z$ basis. The difference between case (ii) and case (iii) is also explained in the same manner, that is, a small $r$ leads to a small S/N ratio in $X$ basis. 
These intuitions are supported by the results in Fig.~\ref{keyrategraph} (b) and Fig.~\ref{keyrategraph2} (b), in which the optimal value of $p_Z$ decreases with fiber length in the passive case while it is fixed to be high in the active case.

\begin{figure*}[t]
 \centering
 \includegraphics[keepaspectratio, scale=0.4]
      {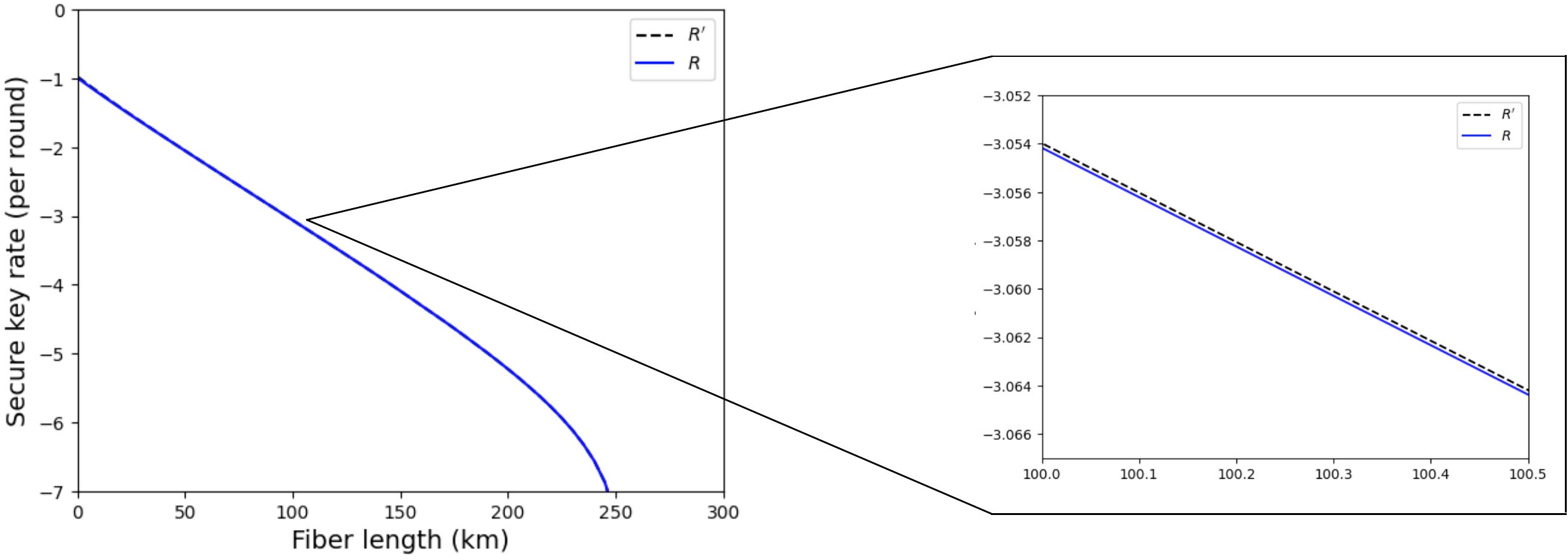}
 \caption{Secure key rate $R$ and virtual rate $R'$ given by Eq. (\ref{virtualkeyrate}) on a logarithmic scale with base 10 as a function of fiber length (km). The blue solid line is the secure key rate $R$ of our result with $r=1$. The black dashed line is the virtual rate $R'$ with $r=1$, in which we assume that all incoming signals contain no more than one photon.  The parameter setting for calculation is the same as that of Fig.~\ref{keyrategraph}. The probability $p_Z$ and the signal intensity $\mu$ are optimized at each distance. 
 }
 \label{tightness}
\end{figure*}

\subsection{Tightness of our bounds}
Although our security proof focuses only on Bob's single-photon detection events and discards multiple-photon events for simplicity, this does not significantly affect the key rate in practice. 
We emphasize that the decoy-state method enables us to obtain the upper bound $\bar{Q}_{\rm cross}^{(n_A=1)}$ on the ratio $Q_{\rm cross}^{(n_A=1)}$ of the number of cross-click rounds with Alice's single-photon emission to the number of non-sampling rounds. 
Since multiple photons do not arrive at Bob's site in rounds with Alice's single-photon emission as long as Eve does not intentionally insert multiple photons, cross-click events are induced by dark counts of detectors. That is, $\bar{Q}_{\rm cross}^{(n_A=1)} = {\rm max} \{O(\eta d),  O(d^2) \}$, which is negligible compared to the lower bound on the sifted-key rate $\barbelow{Q}_Z^{(n_A=1)} = O(\eta)$ where $\eta$ represents the overall transmittance including the channel transmittance and the quantum efficiency. Thus, $\barbelow{Q}_Z^{(n_A=1)} - \alpha \bar{Q}_{\rm cross}^{(n_A =1)}$ in Eq. (\ref{finalkeyrate}) is expected to almost equal $\barbelow{Q}_Z^{(n_A=1)}$.


In order to confirm the discussion here, we consider the virtual rate $R'$ described as  
\begin{align}
R'  \coloneqq  \barbelow{Q}_Z^{(n_A=0)} + \barbelow{Q}_Z^{(n_A=1)}  \left(1- h \left( \frac{p_Z\tilde{p}_Z \bar{E}_X^{(n_A =1)}}{p_X\tilde{p}_X \barbelow{Q}_Z^{(n_A=1)}}  \right)\right) 
- f_{\rm EC}, 
\label{virtualkeyrate}
\end{align}
in which $\alpha \bar{Q}_{\rm cross}^{(n_A=1)}$ is removed from Eq. (\ref{finalkeyrate}). 
This corresponds to an optimistic assumption that all incoming signals to Bob contain no more than one photon. 
With parameters used in Sec. \ref{simulation} and setting $r=1$, we calculated $R$ and $R'$ to compare them as shown in Fig.~\ref{tightness}. The vertical axis is on a logarithmic scale with base 10.
As expected, we can see that the difference between $R$ and $R'$ is negligible.

\section{Conclusion} \label{conclusion}
Here we have proved the security of the decoy-state BB84 protocol with receiver's passive biased basis choice in the asymptotic limit. 
The difficulty is stemming from the fact that the probability of basis choice depends on the incoming photon number. 
We have overcome the problem by estimating the contribution of multiple photons from the ratio of cross-click rounds, to limit our analysis to Bob's single-photon space, combined with the security proof based on complementarity. 
The numerical results show that, thanks to the application of the decoy-state method to cross-click events, the passive protocol achieves a key generation rate comparable to that of the active protocol, except at long distances. Although this work has focused on the security proof in the asymptotic limit, it can be extended to a finite-key analysis using concentration inequalities such as Kato's inequality \cite{2020Kato}. 
On the other hand, since fully passive optical setups are effective in preventing side-channel attacks, a version of the BB84 protocol with passive setups on Alice's side was proposed, and its security was proven \cite{2023Wang, 2023Zapatero, 2024Zapatero}.
In our work, Alice's state preparation is assumed to be performed actively rather than passively.  To move toward fully passive QKD with a biased basis choice, it would be desirable to combine our work with their security proof. However, such integration is non-trivial and is left as an open problem for future work. 

{\it Note added}: After the completion of this work, a related study was presented by Wang {\it et al.} \cite{2025Wang}. Their work can be seen as a generalization of ours, in the sense that it incorporates finite-key effects and considers various device imperfections, such as quantum efficiency mismatch in $Z$ (or $X$) line. However, to the best of our knowledge, our work provides the first analytical security proof of the passive-biased BB84 protocol in the asymptotic limit.

\section*{Acknowledgement}
We thank Akihiro Mizutani, Kiyoshi Tamaki, Takuya Ikuta, Toshimori Honjo, Lars Kamin and Devashish Tupkary for helpful discussions. 
We thank the support of `MIC R\&D of ICT Priority Technology (JPMI00316)' of Ministry of Internal Affairs and Communications of Japan.

\section*{Appendix}
\appendix
\subsection{Theoretical difference between active and passive basis choice} \label{Difference}
Here we show the qualitative difficulty of the passive-biased BB84 protocol from the perspective of security proof. 
We assume that optical loss is balanced between $Z$ line and $X$ line (i.e., $r=1$) and there is no dark count for simplicity. 
Let us denote $Q_X^{(n_A =1)} \coloneqq \sum_{n_B\geq0} Q_X^{(1, n_B)}$.

For the active-biased protocol, 
\be
E_{\rm ph}^{(1, n_B)} \sim \frac{p_Z^2}{p_X^2} E_X^{(1,n_B)}, ~~Q_Z^{(1,n_B)} \sim \frac{p_Z^2}{p_X^2} Q_X^{(1,n_B)}
\ee
hold for any $n_B$ in the asymptotic limit because when $n_A=1$, Alice prepares the state $\ket{\Phi_Z^{(1)}}_{AA'}$ in Eq. (\ref{single-equivalence}) and sends system $A'$ to Bob, followed by their basis choice according to the probability distribution $(p_Z, p_X)$. 
Thus, in the asymptotic limit, the phase error rate $e_{\rm ph}^{(n_A=1)} \coloneqq E_{\rm ph}^{(n_A=1)}/Q_Z^{(n_A=1)} $ for $n_A =1$ equals the observed error rate in $X$ basis; 
\be
e_{\rm ph}^{(n_A=1)} = \frac{E_{\rm ph}^{(n_A=1)}}{Q_Z^{(n_A=1)}} 
= \frac{\sum\limits_{n_B\geq0} E_{\rm ph}^{(1, n_B)}}{\sum\limits_{n_B\geq0} Q_Z^{(1, n_B)}}
\sim \frac{ \frac{p_Z^2}{p_X^2} \sum\limits_{n_B\geq0} E_X^{(1, n_B)}}{\frac{p_Z^2}{p_X^2} \sum\limits_{n_B\geq0} Q_X^{(1, n_B)}}
= \frac{E_X^{(n_A=1)}}{Q_X^{(n_A=1)}}. 
\label{activephaseerror}
\ee

On the other hand, if Bob chooses his basis passively (and assigns no bit to cross-click rounds), Bob's basis is conclusive only when all $n_B$ photons go to $Z$ line or $X$ line.  
The ratio of $Z$ basis to $X$ basis is $p_Z^{n_B} : p_X^{n_B}$ depending on $n_B$. Since Alice actively chooses her basis according to the probability distribution $(p_Z, p_X)$, 
\be
E_{\rm ph}^{(1, n_B)} \sim \frac{p_Z^{n_B+1}}{p_X^{n_B+1}} E_X^{(1,n_B)},~~Q_Z^{(1,n_B)} \sim \frac{p_Z^{n_B+1}}{p_X^{n_B+1}} Q_X^{(1,n_B)}
\label{simpleanalysis}
\ee
hold in the asymptotic limit. The phase error rate for rounds with $n_A =1$ is 
\be
e_{\rm ph}^{(n_A=1)} = \frac{E_{\rm ph}^{(n_A=1)}}{Q_Z^{(n_A=1)}} 
= \frac{\sum\limits_{n_B\geq0} E_{\rm ph}^{(1, n_B)}}{\sum\limits_{n_B\geq0} Q_Z^{(1, n_B)}}
\sim \frac{ \sum\limits_{n_B\geq0} \frac{p_Z^{n_B+1}}{p_X^{n_B+1}} E_X^{(1, n_B)}}{\sum\limits_{n_B\geq0} \frac{p_Z^{n_B+1}}{p_X^{n_B+1}} Q_X^{(1, n_B)}},
\ee
which is not identical to the $X$-basis error rate $E_X^{(n_A=1)} / Q_X^{(n_A=1)}$ for $n_A =1$ except the case of balanced basis choice, i.e., $p_Z=p_X=1/2$. 
As a consequence, if we consider the sum of all $n_B$ mentioned in the above scenario, the relation Eq. (\ref{activephaseerror}) between the phase error and the $X$-basis error which is satisfied in the active protocol does not hold. This motivates us to analyze the security with fixed $n_B$ to accomplish a simple proof.

\subsection{Decoy state analysis for receiver's passive setup}\label{decoyproof}
Here we derive bounds on $Q_Z^{(n_A=0)}, Q_Z^{(n_A=1)}, E_X^{(n_A=1)}$ and $Q_{\rm cross}^{(n_A=1)}$ with decoy-state method using weak coherent states and the vacuum. 
In a similar way to \cite{2005Ma}, we first derive those bounds with intensities $\mu_A \in \{ \mu, \nu_1, \nu_2 \}$ satisfying $0 \leq \nu_2 < \nu_1$ and $\nu_1 + \nu_2 < \mu$, followed by taking $\nu_2= 0$ at last. 
Let $Y_{Z|n'_A}$, $Y_{X|n'_A}^{\rm error}$ and $Y_{{\rm cross}|n'_A}$ be 
\begin{equation}
\begin{split}
Y_{Z|n'_A} &\coloneqq \frac{m(y_B =1 \land W_A = W_B =Z \land n_A = n'_A)}{m(n_A = n'_A)},  \\
Y_{X|n'_A}^{\rm error} &\coloneqq \frac{m(y_B =1 \land W_A = W_B =X \land n_A = n'_A \land b_A \neq b_B)}{m(n_A = n'_A)},  \\
Y_{{\rm cross}|n'_A} &\coloneqq \frac{m(y_B =1 \land  W_B =\bot \land n_A = n'_A)}{m(n_A = n'_A)},
\end{split}
\end{equation}
respectively.
For $\mu_A \in \{ \mu, \nu_1, \nu_2 \}$, the following relations hold:
\begin{gather}
Q_{Z|\mu_A} \sim \sum\limits_{n_A\geq0} Y_{Z|n_A} \frac{\mu_A^{n_A}}{n_A !} e^{-\mu_A}, \label{Qdecoy} \\
E_{X|\mu_A} \sim \sum\limits_{n_A\geq0} Y_{X|n_A}^{\rm error} \frac{\mu_A^{n_A}}{n_A !} e^{-\mu_A}, \label{Edecoy}\\
Q_{{\rm cross}|\mu_A} \sim \sum\limits_{n_A\geq0} Y_{{\rm cross}|n_A} \frac{\mu_A^{n_A}}{n_A !} e^{-\mu_A}. \label{Cdecoy}
\end{gather}
From Eq. (\ref{Qdecoy}) with $\mu_A = \nu_1$ and $\mu_A= \nu_2$, we obtain 
\begin{align}
\nu_1 Q_{Z|\nu_2} e^{\nu_2} - \nu_2 Q_{Z|\nu_1} e^{\nu_1}
\sim (\nu_1-\nu_2) Y_{Z|0}  - \nu_1\nu_2 \sum\limits_{n_A\geq2} Y_{Z|n_A} \frac{\nu_1^{n_A-1} - \nu_2^{n_A-1}}{n_A!} 
\leq (\nu_1-\nu_2) Y_{Z|0}. 
\end{align}
A lower bound on $Y_{Z|0}$ is given by 
\be
Y_{Z|0} \gtrsim Y_{Z|0}^{\rm low} \coloneqq {\rm max} \left \{ \frac{\nu_1Q_{Z|\nu_2} e^{\nu_2} - \nu_2Q_{Z|\nu_1} e^{\nu_1}}{\nu_1-\nu_2}, 0 \right\}, 
\ee
which leads to 
\begin{equation}
\begin{split}
Q_Z^{(n_A=0)} &\sim (p_{\mu}P_\mu^{\rm Poisson}(0) + p_{\nu_1} P_{\nu_1}^{\rm Poisson}(0) + p_{\nu_2} P_{\nu_2}^{\rm Poisson}(0)) Y_{Z|0}  \\
& \gtrsim (p_{\mu}e^{-\mu} + p_{\nu_1}e^{-\nu_1} + p_{\nu_2}e^{-\nu_2}) Y_{Z|0}^{\rm low}.
\label{decoy0}
\end{split}
\end{equation}
Next, from Eq. (\ref{Qdecoy}) with $\mu_A = \mu$, 
\be
Q_{Z|\mu} e^{\mu} \sim \sum\limits_{n_A\geq0} Y_{Z|n_A} \frac{\mu^{n_A}}{n_A!}
\ee
holds. 
The contribution of more than one photon is given by
\be
\sum\limits_{n_A\geq2} Y_{Z|n_A} \frac{\mu^{n_A}}{n_A!} \sim Q_{Z|\mu} e^{\mu} - Y_{Z|0} - Y_{Z|1} \mu.
\label{contrimorethanone}
\ee
We obtain the following inequality from Eq. (\ref{Qdecoy}) with $\mu_A= \nu_1$ and  $\mu_A= \nu_2$;
\begin{equation}
\begin{split}
&Q_{Z|\nu_1} e^{\nu_1} - Q_{Z|\nu_2} e^{\nu_2}  \\
&\sim Y_{Z|1}(\nu_1 - \nu_2) + \sum\limits_{n_A\geq2}  \frac{Y_{Z|n_A}}{n_A!}(\nu_1^{n_A} - \nu_2^{n_A})  
= Y_{Z|1}(\nu_1 - \nu_2) + \sum\limits_{n_A\geq2}  \frac{Y_{Z|n_A}}{n_A!} \mu^{n_A} \left( \left(\frac{\nu_1}{\mu} \right)^{n_A} - \left(\frac{\nu_2}{\mu} \right)^{n_A} \right)  \\
&\leq Y_{Z|1}(\nu_1 - \nu_2) + \frac{\nu_1^2 - \nu_2^2}{\mu^2}  \sum\limits_{n_A\geq2} \frac{Y_{Z|n_A}}{n_A!} \mu^{n_A} 
\sim Y_{Z|1}(\nu_1 - \nu_2) + \frac{\nu_1^2 - \nu_2^2}{\mu^2} (Q_{Z|\mu} e^{\mu} - Y_{Z|0} - Y_{Z|1} \mu)  \\
&\lesssim Y_{Z|1}(\nu_1 - \nu_2) + \frac{\nu_1^2 - \nu_2^2}{\mu^2} (Q_{Z|\mu} e^{\mu} - Y_{Z|0}^{\rm low} - Y_{Z|1} \mu)  \\
&=\left(\nu_1 - \nu_2 - \frac{\nu_1^2 - \nu_2^2}{\mu} \right) Y_{Z|1} + \frac{\nu_1^2 - \nu_2^2}{\mu^2} (Q_{Z|\mu} e^{\mu} - Y_{Z|0}^{\rm low}),
\end{split}
\end{equation}
where the first inequality is satisfied because $a^i - b^i \leq a^2 - b^2$ holds for $i\geq 2$ if $0<a+b<1$ and $\nu_1 + \nu_2  < \mu$ are assumed, and the third equality holds from Eq. (\ref{contrimorethanone}).  
Thus a lower bound on $Y_{Z|1}$ is given by 
\begin{align}
&Y_{Z|1} \gtrsim \frac{\mu}{\mu \nu_1 - \mu \nu_2 - \nu_1^2 + \nu_2 ^2} \left( Q_{Z|\nu_1} e^{\nu_1} - Q_{Z|\nu_2} e^{\nu_2} - \frac{\nu_1^2 - \nu_2^2}{\mu^2} 
(Q_{Z|\mu} e^{\mu} - Y_{Z|0}^{\rm low})
\right),
\end{align}
which leads to 
\begin{equation}
\begin{split}
Q_Z^{(n_A=1)} &\sim (p_{\mu}P_\mu^{\rm Poisson}(1) + p_{\nu_1} P_{\nu_1}^{\rm Poisson}(1) + p_{\nu_2} P_{\nu_2}^{\rm Poisson}(1)) Y_{Z|1}  \\
&\gtrsim
\frac{\mu(p_{\mu} \mu e^{-\mu} + p_{\nu_1} \nu_1 e^{-\nu_1} + p_{\nu_2} \nu_2 e^{-\nu_2} )}{\mu \nu_1 - \mu \nu_2 - \nu_1^2 + \nu_2 ^2} \left( Q_{Z|\nu_1} e^{\nu_1} - Q_{Z|\nu_2} e^{\nu_2} - \frac{\nu_1^2 - \nu_2^2}{\mu^2} 
(Q_{Z|\mu} e^{\mu} - Y_{Z|0}^{\rm low}) \right).
\label{Qlower}
\end{split}
\end{equation}
For the error ratio, from Eq. (\ref{Edecoy}) with $\mu_A = \nu_1$ and $\mu_A = \nu_2$, we obtain 
\begin{equation}
\begin{split}
E_{X|\nu_1} e^{\nu_1} \sim Y_{X|0}^{\rm error} + Y_{X|1}^{\rm error} \nu_1 + \sum\limits_{n_A\geq2} Y_{X|n_A}^{\rm error} \frac{\nu_1^{n_A}}{n_A!}, \\
E_{X|\nu_2} e^{\nu_2} \sim Y_{X|0}^{\rm error} + Y_{X|1}^{\rm error} \nu_2 + \sum\limits_{n_A\geq2} Y_{X|n_A}^{\rm error} \frac{\nu_2^{n_A}}{n_A!}.
\end{split}
\end{equation}
Since $\nu_1 > \nu_2$ is satisfied, 
\be
Y_{X|1}^{\rm error} \lesssim \frac{E_{X|\nu_1} e^{\nu_1} - E_{X|\nu_2} e^{\nu_2}}{\nu_1 - \nu_2}
\ee
holds, which leads to 
\begin{equation}
\begin{split}
E_X^{(n_A=1)} 
&\sim (p_{\mu}P_\mu^{\rm Poisson}(1) + p_{\nu_1} P_{\nu_1}^{\rm Poisson}(1) + p_{\nu_2} P_{\nu_2}^{\rm Poisson}(1)) Y_{Z|1}^{\rm error}  \\
&\lesssim (p_{\mu} \mu e^{-\mu} + p_{\nu_1} \nu_1 e^{-\nu_1} + p_{\nu_2} \nu_2 e^{-\nu_2}) \frac{E_{X|\nu_1} e^{\nu_1} - E_{X|\nu_2} e^{\nu_2}}{\nu_1 - \nu_2}. 
\label{Eupper}
\end{split}
\end{equation}
For the cross-click ratio, from Eq. (\ref{Cdecoy}) with $\mu_A = \nu_1$ and $\mu_A = \nu_2$, we obtain 
\begin{equation}
\begin{split}
Q_{{\rm cross}|\nu_1} e^{\nu_1} \sim Y_{{\rm cross}|0} + Y_{{\rm cross}|1} \nu_1 + \sum\limits_{n_A\geq2} Y_{{\rm cross}|n_A} \frac{\nu_1^{n_A}}{n_A!}, \\
Q_{{\rm cross}|\nu_2} e^{\nu_2} \sim Y_{{\rm cross}|0} + Y_{{\rm cross}|1} \nu_2 + \sum\limits_{n_A\geq2} Y_{{\rm cross}|n_A} \frac{\nu_2^{n_A}}{n_A!}.
\end{split}
\end{equation}
Since $\nu_1 > \nu_2$ is satisfied, 
\be
Y_{{\rm cross}|1} \lesssim \frac{Q_{{\rm cross}|\nu_1} e^{\nu_1} - Q_{{\rm cross}|\nu_2} e^{\nu_2}}{\nu_1 - \nu_2}
\ee
holds, which leads to 
\begin{equation}
\begin{split}
Q_{\rm cross}^{(n_A=1)} 
&\sim (p_{\mu}P_\mu^{\rm Poisson}(1) + p_{\nu_1} P_{\nu_1}^{\rm Poisson}(1) + p_{\nu_2} P_{\nu_2}^{\rm Poisson}(1)) Y_{{\rm cross}|1}  \\
&\lesssim (p_{\mu} \mu e^{-\mu} + p_{\nu_1} \nu_1 e^{-\nu_1} + p_{\nu_2} \nu_2 e^{-\nu_2}) \frac{Q_{{\rm cross}|\nu_1} e^{\nu_1} - Q_{{\rm cross}|\nu_2} e^{\nu_2}}{\nu_1 - \nu_2}.
\label{Cupper}
\end{split}
\end{equation}
In Eqs. (\ref{decoy0}), (\ref{Qlower}), (\ref{Eupper}) and (\ref{Cupper}), by taking $\nu_1 = \nu$ and $\nu_2 = 0$, we obtain Eq. (\ref{decoybounds}). 


\bibliography{kawakamibibD}

\begin{thebibliography}{35}%
\makeatletter
\providecommand \@ifxundefined [1]{%
 \@ifx{#1\undefined}
}%
\providecommand \@ifnum [1]{%
 \ifnum #1\expandafter \@firstoftwo
 \else \expandafter \@secondoftwo
 \fi
}%
\providecommand \@ifx [1]{%
 \ifx #1\expandafter \@firstoftwo
 \else \expandafter \@secondoftwo
 \fi
}%
\providecommand \natexlab [1]{#1}%
\providecommand \enquote  [1]{``#1''}%
\providecommand \bibnamefont  [1]{#1}%
\providecommand \bibfnamefont [1]{#1}%
\providecommand \citenamefont [1]{#1}%
\providecommand \href@noop [0]{\@secondoftwo}%
\providecommand \href [0]{\begingroup \@sanitize@url \@href}%
\providecommand \@href[1]{\@@startlink{#1}\@@href}%
\providecommand \@@href[1]{\endgroup#1\@@endlink}%
\providecommand \@sanitize@url [0]{\catcode `\\12\catcode `\$12\catcode
  `\&12\catcode `\#12\catcode `\^12\catcode `\_12\catcode `\%12\relax}%
\providecommand \@@startlink[1]{}%
\providecommand \@@endlink[0]{}%
\providecommand \url  [0]{\begingroup\@sanitize@url \@url }%
\providecommand \@url [1]{\endgroup\@href {#1}{\urlprefix }}%
\providecommand \urlprefix  [0]{URL }%
\providecommand \Eprint [0]{\href }%
\providecommand \doibase [0]{http://dx.doi.org/}%
\providecommand \selectlanguage [0]{\@gobble}%
\providecommand \bibinfo  [0]{\@secondoftwo}%
\providecommand \bibfield  [0]{\@secondoftwo}%
\providecommand \translation [1]{[#1]}%
\providecommand \BibitemOpen [0]{}%
\providecommand \bibitemStop [0]{}%
\providecommand \bibitemNoStop [0]{.\EOS\space}%
\providecommand \EOS [0]{\spacefactor3000\relax}%
\providecommand \BibitemShut  [1]{\csname bibitem#1\endcsname}%
\let\auto@bib@innerbib\@empty
\bibitem [{\citenamefont {Bennett}\ and\ \citenamefont
  {Brassard}(1984)}]{1984Bennett}%
  \BibitemOpen
  \bibfield  {author} {\bibinfo {author} {\bibfnamefont {C.~H.}\ \bibnamefont
  {Bennett}}\ and\ \bibinfo {author} {\bibfnamefont {G.}~\bibnamefont
  {Brassard}},\ }in\ \href@noop {} {\emph {\bibinfo {booktitle} {Proceedings of
  IEEE International Conference on Computers, Systems and Signal
  Processing}}},\ Vol.\ \bibinfo {volume} {175},\ \bibinfo {organization}
  {Bangalore, India}\ (\bibinfo  {publisher} {IEEE Press},\ \bibinfo {address}
  {New York},\ \bibinfo {year} {1984})\BibitemShut {NoStop}%
\bibitem [{\citenamefont {Hwang}(2003)}]{2003Hwang}%
  \BibitemOpen
  \bibfield  {author} {\bibinfo {author} {\bibfnamefont {W.-Y.}\ \bibnamefont
  {Hwang}},\ }\href {\doibase 10.1103/PhysRevLett.91.057901} {\bibfield
  {journal} {\bibinfo  {journal} {Phys. Rev. Lett.}\ }\textbf {\bibinfo
  {volume} {91}},\ \bibinfo {pages} {057901} (\bibinfo {year}
  {2003})}\BibitemShut {NoStop}%
\bibitem [{\citenamefont {Wang}(2005{\natexlab{a}})}]{2005WangPRL}%
  \BibitemOpen
  \bibfield  {author} {\bibinfo {author} {\bibfnamefont {X.-B.}\ \bibnamefont
  {Wang}},\ }\href {\doibase 10.1103/PhysRevLett.94.230503} {\bibfield
  {journal} {\bibinfo  {journal} {Phys. Rev. Lett.}\ }\textbf {\bibinfo
  {volume} {94}},\ \bibinfo {pages} {230503} (\bibinfo {year}
  {2005}{\natexlab{a}})}\BibitemShut {NoStop}%
\bibitem [{\citenamefont {Lo}\ \emph {et~al.}(2005)\citenamefont {Lo},
  \citenamefont {Ma},\ and\ \citenamefont {Chen}}]{2005Lo}%
  \BibitemOpen
  \bibfield  {author} {\bibinfo {author} {\bibfnamefont {H.-K.}\ \bibnamefont
  {Lo}}, \bibinfo {author} {\bibfnamefont {X.}~\bibnamefont {Ma}}, \ and\
  \bibinfo {author} {\bibfnamefont {K.}~\bibnamefont {Chen}},\ }\href {\doibase
  10.1103/PhysRevLett.94.230504} {\bibfield  {journal} {\bibinfo  {journal}
  {Phys. Rev. Lett.}\ }\textbf {\bibinfo {volume} {94}},\ \bibinfo {pages}
  {230504} (\bibinfo {year} {2005})}\BibitemShut {NoStop}%
\bibitem [{\citenamefont {Wang}(2005{\natexlab{b}})}]{2005Wang}%
  \BibitemOpen
  \bibfield  {author} {\bibinfo {author} {\bibfnamefont {X.-B.}\ \bibnamefont
  {Wang}},\ }\href {\doibase 10.1103/PhysRevA.72.012322} {\bibfield  {journal}
  {\bibinfo  {journal} {Phys. Rev. A}\ }\textbf {\bibinfo {volume} {72}},\
  \bibinfo {pages} {012322} (\bibinfo {year} {2005}{\natexlab{b}})}\BibitemShut
  {NoStop}%
\bibitem [{\citenamefont {Beaudry}\ \emph {et~al.}(2008)\citenamefont
  {Beaudry}, \citenamefont {Moroder},\ and\ \citenamefont
  {L\"utkenhaus}}]{2008Lutkenhaus}%
  \BibitemOpen
  \bibfield  {author} {\bibinfo {author} {\bibfnamefont {N.~J.}\ \bibnamefont
  {Beaudry}}, \bibinfo {author} {\bibfnamefont {T.}~\bibnamefont {Moroder}}, \
  and\ \bibinfo {author} {\bibfnamefont {N.}~\bibnamefont {L\"utkenhaus}},\
  }\href {\doibase 10.1103/PhysRevLett.101.093601} {\bibfield  {journal}
  {\bibinfo  {journal} {Phys. Rev. Lett.}\ }\textbf {\bibinfo {volume} {101}},\
  \bibinfo {pages} {093601} (\bibinfo {year} {2008})}\BibitemShut {NoStop}%
\bibitem [{\citenamefont {Tsurumaru}\ and\ \citenamefont
  {Tamaki}(2008)}]{2008Tsurumaru}%
  \BibitemOpen
  \bibfield  {author} {\bibinfo {author} {\bibfnamefont {T.}~\bibnamefont
  {Tsurumaru}}\ and\ \bibinfo {author} {\bibfnamefont {K.}~\bibnamefont
  {Tamaki}},\ }\href {\doibase 10.1103/PhysRevA.78.032302} {\bibfield
  {journal} {\bibinfo  {journal} {Phys. Rev. A}\ }\textbf {\bibinfo {volume}
  {78}},\ \bibinfo {pages} {032302} (\bibinfo {year} {2008})}\BibitemShut
  {NoStop}%
\bibitem [{\citenamefont {Koashi}(2006)}]{2006Koashi}%
  \BibitemOpen
  \bibfield  {author} {\bibinfo {author} {\bibfnamefont {M.}~\bibnamefont
  {Koashi}},\ }\href {https://arxiv.org/abs/quant-ph/0609180} {\bibfield
  {journal} {\bibinfo  {journal} {\rm arXiv:quant-ph/0609180}\ } (\bibinfo
  {year} {2006})}\BibitemShut {NoStop}%
\bibitem [{\citenamefont {Koashi}(2009)}]{2009Koashi}%
  \BibitemOpen
  \bibfield  {author} {\bibinfo {author} {\bibfnamefont {M.}~\bibnamefont
  {Koashi}},\ }\href
  {https://iopscience.iop.org/article/10.1088/1367-2630/11/4/045018} {\bibfield
   {journal} {\bibinfo  {journal} {New J. Phys.}\ }\textbf {\bibinfo {volume}
  {11}},\ \bibinfo {pages} {045018} (\bibinfo {year} {2009})}\BibitemShut
  {NoStop}%
\bibitem [{\citenamefont {Tomamichel}\ and\ \citenamefont
  {Renner}(2011)}]{2011Tomamichel}%
  \BibitemOpen
  \bibfield  {author} {\bibinfo {author} {\bibfnamefont {M.}~\bibnamefont
  {Tomamichel}}\ and\ \bibinfo {author} {\bibfnamefont {R.}~\bibnamefont
  {Renner}},\ }\href {\doibase 10.1103/PhysRevLett.106.110506} {\bibfield
  {journal} {\bibinfo  {journal} {Phys. Rev. Lett.}\ }\textbf {\bibinfo
  {volume} {106}},\ \bibinfo {pages} {110506} (\bibinfo {year}
  {2011})}\BibitemShut {NoStop}%
\bibitem [{\citenamefont {Lo}\ \emph {et~al.}(2004)\citenamefont {Lo},
  \citenamefont {Chau},\ and\ \citenamefont {Ardehali}}]{2004Lo}%
  \BibitemOpen
  \bibfield  {author} {\bibinfo {author} {\bibfnamefont {H.-K.}\ \bibnamefont
  {Lo}}, \bibinfo {author} {\bibfnamefont {H.}~\bibnamefont {Chau}}, \ and\
  \bibinfo {author} {\bibfnamefont {M.}~\bibnamefont {Ardehali}},\ }\href@noop
  {} {\bibfield  {journal} {\bibinfo  {journal} {Journal of Cryptology}\
  }\textbf {\bibinfo {volume} {18(2)}},\ \bibinfo {pages} {133} (\bibinfo
  {year} {2004})}\BibitemShut {NoStop}%
\bibitem [{\citenamefont {Liao}\ \emph {et~al.}(2017)\citenamefont {Liao},
  \citenamefont {Cai}, \citenamefont {Liu}, \citenamefont {Zhang},
  \citenamefont {Li}, \citenamefont {Ren}, \citenamefont {Yin}, \citenamefont
  {Shen}, \citenamefont {Cao}, \citenamefont {Li}, \citenamefont {Li},
  \citenamefont {Chen}, \citenamefont {Sun}, \citenamefont {Jia}, \citenamefont
  {Wu}, \citenamefont {Jiang}, \citenamefont {Wang}, \citenamefont {Huang},
  \citenamefont {Wang}, \citenamefont {Zhou}, \citenamefont {Deng},
  \citenamefont {Xi}, \citenamefont {Ma}, \citenamefont {Hu}, \citenamefont
  {Zhang}, \citenamefont {Chen}, \citenamefont {Liu}, \citenamefont {Wang},
  \citenamefont {Zhu}, \citenamefont {Lu}, \citenamefont {Shu}, \citenamefont
  {Peng}, \citenamefont {Wang},\ and\ \citenamefont {Pan}}]{2017Liao}%
  \BibitemOpen
  \bibfield  {author} {\bibinfo {author} {\bibfnamefont {S.-K.}\ \bibnamefont
  {Liao}}, \bibinfo {author} {\bibfnamefont {W.-Q.}\ \bibnamefont {Cai}},
  \bibinfo {author} {\bibfnamefont {W.-Y.}\ \bibnamefont {Liu}}, \bibinfo
  {author} {\bibfnamefont {L.}~\bibnamefont {Zhang}}, \bibinfo {author}
  {\bibfnamefont {Y.}~\bibnamefont {Li}}, \bibinfo {author} {\bibfnamefont
  {J.-G.}\ \bibnamefont {Ren}}, \bibinfo {author} {\bibfnamefont
  {J.}~\bibnamefont {Yin}}, \bibinfo {author} {\bibfnamefont {Q.}~\bibnamefont
  {Shen}}, \bibinfo {author} {\bibfnamefont {Y.}~\bibnamefont {Cao}}, \bibinfo
  {author} {\bibfnamefont {Z.-P.}\ \bibnamefont {Li}}, \bibinfo {author}
  {\bibfnamefont {F.-Z.}\ \bibnamefont {Li}}, \bibinfo {author} {\bibfnamefont
  {X.-W.}\ \bibnamefont {Chen}}, \bibinfo {author} {\bibfnamefont {L.-H.}\
  \bibnamefont {Sun}}, \bibinfo {author} {\bibfnamefont {J.-J.}\ \bibnamefont
  {Jia}}, \bibinfo {author} {\bibfnamefont {J.-C.}\ \bibnamefont {Wu}},
  \bibinfo {author} {\bibfnamefont {X.-J.}\ \bibnamefont {Jiang}}, \bibinfo
  {author} {\bibfnamefont {J.-F.}\ \bibnamefont {Wang}}, \bibinfo {author}
  {\bibfnamefont {Y.-M.}\ \bibnamefont {Huang}}, \bibinfo {author}
  {\bibfnamefont {Q.}~\bibnamefont {Wang}}, \bibinfo {author} {\bibfnamefont
  {Y.-L.}\ \bibnamefont {Zhou}}, \bibinfo {author} {\bibfnamefont
  {L.}~\bibnamefont {Deng}}, \bibinfo {author} {\bibfnamefont {T.}~\bibnamefont
  {Xi}}, \bibinfo {author} {\bibfnamefont {L.}~\bibnamefont {Ma}}, \bibinfo
  {author} {\bibfnamefont {T.}~\bibnamefont {Hu}}, \bibinfo {author}
  {\bibfnamefont {Q.}~\bibnamefont {Zhang}}, \bibinfo {author} {\bibfnamefont
  {Y.-A.}\ \bibnamefont {Chen}}, \bibinfo {author} {\bibfnamefont {N.-L.}\
  \bibnamefont {Liu}}, \bibinfo {author} {\bibfnamefont {X.-B.}\ \bibnamefont
  {Wang}}, \bibinfo {author} {\bibfnamefont {Z.-C.}\ \bibnamefont {Zhu}},
  \bibinfo {author} {\bibfnamefont {C.-Y.}\ \bibnamefont {Lu}}, \bibinfo
  {author} {\bibfnamefont {R.}~\bibnamefont {Shu}}, \bibinfo {author}
  {\bibfnamefont {C.-Z.}\ \bibnamefont {Peng}}, \bibinfo {author}
  {\bibfnamefont {J.-Y.}\ \bibnamefont {Wang}}, \ and\ \bibinfo {author}
  {\bibfnamefont {J.-W.}\ \bibnamefont {Pan}},\ }\href {\doibase
  10.1038/nature23655} {\bibfield  {journal} {\bibinfo  {journal} {Nature}\
  }\textbf {\bibinfo {volume} {549}},\ \bibinfo {pages} {43} (\bibinfo {year}
  {2017})}\BibitemShut {NoStop}%
\bibitem [{\citenamefont {Sivasankaran}\ \emph {et~al.}(2022)\citenamefont
  {Sivasankaran}, \citenamefont {Liu}, \citenamefont {Mihm},\ and\
  \citenamefont {Ling}}]{2022Sivasankaran}%
  \BibitemOpen
  \bibfield  {author} {\bibinfo {author} {\bibfnamefont {S.}~\bibnamefont
  {Sivasankaran}}, \bibinfo {author} {\bibfnamefont {C.}~\bibnamefont {Liu}},
  \bibinfo {author} {\bibfnamefont {M.}~\bibnamefont {Mihm}}, \ and\ \bibinfo
  {author} {\bibfnamefont {A.}~\bibnamefont {Ling}},\ }in\ \href {\doibase
  10.1109/ICSOS53063.2022.9749724} {\emph {\bibinfo {booktitle} {2022 IEEE
  International Conference on Space Optical Systems and Applications
  (ICSOS)}}}\ (\bibinfo {year} {2022})\ pp.\ \bibinfo {pages}
  {51--56}\BibitemShut {NoStop}%
\bibitem [{\citenamefont {Roger}\ \emph {et~al.}(2023)\citenamefont {Roger},
  \citenamefont {Singh}, \citenamefont {Perumangatt}, \citenamefont {Marangon},
  \citenamefont {Sanzaro}, \citenamefont {Smith}, \citenamefont {Woodward},\
  and\ \citenamefont {Shields}}]{2023Shields}%
  \BibitemOpen
  \bibfield  {author} {\bibinfo {author} {\bibfnamefont {T.}~\bibnamefont
  {Roger}}, \bibinfo {author} {\bibfnamefont {R.}~\bibnamefont {Singh}},
  \bibinfo {author} {\bibfnamefont {C.}~\bibnamefont {Perumangatt}}, \bibinfo
  {author} {\bibfnamefont {D.~G.}\ \bibnamefont {Marangon}}, \bibinfo {author}
  {\bibfnamefont {M.}~\bibnamefont {Sanzaro}}, \bibinfo {author} {\bibfnamefont
  {P.~R.}\ \bibnamefont {Smith}}, \bibinfo {author} {\bibfnamefont {R.~I.}\
  \bibnamefont {Woodward}}, \ and\ \bibinfo {author} {\bibfnamefont {A.~J.}\
  \bibnamefont {Shields}},\ }\href {\doibase 10.1126/sciadv.adj5873} {\bibfield
   {journal} {\bibinfo  {journal} {Science Advances}\ }\textbf {\bibinfo
  {volume} {9}},\ \bibinfo {pages} {eadj5873} (\bibinfo {year}
  {2023})}\BibitemShut {NoStop}%
\bibitem [{\citenamefont {Li}\ \emph {et~al.}(2025)\citenamefont {Li},
  \citenamefont {Cai}, \citenamefont {Ren}, \citenamefont {Wang}, \citenamefont
  {Yang}, \citenamefont {Zhang}, \citenamefont {Wu}, \citenamefont {Chang},
  \citenamefont {Wu}, \citenamefont {Jin}, \citenamefont {Xue}, \citenamefont
  {Li}, \citenamefont {Liu}, \citenamefont {Yu}, \citenamefont {Tao},
  \citenamefont {Chen}, \citenamefont {Liu}, \citenamefont {Luo}, \citenamefont
  {Zhou}, \citenamefont {Yong}, \citenamefont {Li}, \citenamefont {Li},
  \citenamefont {Jiang}, \citenamefont {Chen}, \citenamefont {Wu},
  \citenamefont {Tong}, \citenamefont {Xie}, \citenamefont {Zhou},
  \citenamefont {Liu}, \citenamefont {Ismail}, \citenamefont {Petruccione},
  \citenamefont {Liu}, \citenamefont {Li}, \citenamefont {Xu}, \citenamefont
  {Cao}, \citenamefont {Yin}, \citenamefont {Shu}, \citenamefont {Wang},
  \citenamefont {Zhang}, \citenamefont {Wang}, \citenamefont {Liao},
  \citenamefont {Peng},\ and\ \citenamefont {Pan}}]{2025Yang}%
  \BibitemOpen
  \bibfield  {author} {\bibinfo {author} {\bibfnamefont {Y.}~\bibnamefont
  {Li}}, \bibinfo {author} {\bibfnamefont {W.-Q.}\ \bibnamefont {Cai}},
  \bibinfo {author} {\bibfnamefont {J.-G.}\ \bibnamefont {Ren}}, \bibinfo
  {author} {\bibfnamefont {C.-Z.}\ \bibnamefont {Wang}}, \bibinfo {author}
  {\bibfnamefont {M.}~\bibnamefont {Yang}}, \bibinfo {author} {\bibfnamefont
  {L.}~\bibnamefont {Zhang}}, \bibinfo {author} {\bibfnamefont {H.-Y.}\
  \bibnamefont {Wu}}, \bibinfo {author} {\bibfnamefont {L.}~\bibnamefont
  {Chang}}, \bibinfo {author} {\bibfnamefont {J.-C.}\ \bibnamefont {Wu}},
  \bibinfo {author} {\bibfnamefont {B.}~\bibnamefont {Jin}}, \bibinfo {author}
  {\bibfnamefont {H.-J.}\ \bibnamefont {Xue}}, \bibinfo {author} {\bibfnamefont
  {X.-J.}\ \bibnamefont {Li}}, \bibinfo {author} {\bibfnamefont
  {H.}~\bibnamefont {Liu}}, \bibinfo {author} {\bibfnamefont {G.-W.}\
  \bibnamefont {Yu}}, \bibinfo {author} {\bibfnamefont {X.-Y.}\ \bibnamefont
  {Tao}}, \bibinfo {author} {\bibfnamefont {T.}~\bibnamefont {Chen}}, \bibinfo
  {author} {\bibfnamefont {C.-F.}\ \bibnamefont {Liu}}, \bibinfo {author}
  {\bibfnamefont {W.-B.}\ \bibnamefont {Luo}}, \bibinfo {author} {\bibfnamefont
  {J.}~\bibnamefont {Zhou}}, \bibinfo {author} {\bibfnamefont {H.-L.}\
  \bibnamefont {Yong}}, \bibinfo {author} {\bibfnamefont {Y.-H.}\ \bibnamefont
  {Li}}, \bibinfo {author} {\bibfnamefont {F.-Z.}\ \bibnamefont {Li}}, \bibinfo
  {author} {\bibfnamefont {C.}~\bibnamefont {Jiang}}, \bibinfo {author}
  {\bibfnamefont {H.-Z.}\ \bibnamefont {Chen}}, \bibinfo {author}
  {\bibfnamefont {C.}~\bibnamefont {Wu}}, \bibinfo {author} {\bibfnamefont
  {X.-H.}\ \bibnamefont {Tong}}, \bibinfo {author} {\bibfnamefont {S.-J.}\
  \bibnamefont {Xie}}, \bibinfo {author} {\bibfnamefont {F.}~\bibnamefont
  {Zhou}}, \bibinfo {author} {\bibfnamefont {W.-Y.}\ \bibnamefont {Liu}},
  \bibinfo {author} {\bibfnamefont {Y.}~\bibnamefont {Ismail}}, \bibinfo
  {author} {\bibfnamefont {F.}~\bibnamefont {Petruccione}}, \bibinfo {author}
  {\bibfnamefont {N.-L.}\ \bibnamefont {Liu}}, \bibinfo {author} {\bibfnamefont
  {L.}~\bibnamefont {Li}}, \bibinfo {author} {\bibfnamefont {F.}~\bibnamefont
  {Xu}}, \bibinfo {author} {\bibfnamefont {Y.}~\bibnamefont {Cao}}, \bibinfo
  {author} {\bibfnamefont {J.}~\bibnamefont {Yin}}, \bibinfo {author}
  {\bibfnamefont {R.}~\bibnamefont {Shu}}, \bibinfo {author} {\bibfnamefont
  {X.-B.}\ \bibnamefont {Wang}}, \bibinfo {author} {\bibfnamefont
  {Q.}~\bibnamefont {Zhang}}, \bibinfo {author} {\bibfnamefont {J.-Y.}\
  \bibnamefont {Wang}}, \bibinfo {author} {\bibfnamefont {S.-K.}\ \bibnamefont
  {Liao}}, \bibinfo {author} {\bibfnamefont {C.-Z.}\ \bibnamefont {Peng}}, \
  and\ \bibinfo {author} {\bibfnamefont {J.-W.}\ \bibnamefont {Pan}},\ }\href
  {\doibase 10.1038/s41586-025-08739-z} {\bibfield  {journal} {\bibinfo
  {journal} {Nature}\ }\textbf {\bibinfo {volume} {640}},\ \bibinfo {pages}
  {47} (\bibinfo {year} {2025})}\BibitemShut {NoStop}%
\bibitem [{\citenamefont {Yin}\ \emph {et~al.}(2020)\citenamefont {Yin},
  \citenamefont {Liu}, \citenamefont {Dai}, \citenamefont {Ci}, \citenamefont
  {Gu}, \citenamefont {Gao}, \citenamefont {Wang},\ and\ \citenamefont
  {Shen}}]{2020Yin}%
  \BibitemOpen
  \bibfield  {author} {\bibinfo {author} {\bibfnamefont {H.-L.}\ \bibnamefont
  {Yin}}, \bibinfo {author} {\bibfnamefont {P.}~\bibnamefont {Liu}}, \bibinfo
  {author} {\bibfnamefont {W.-W.}\ \bibnamefont {Dai}}, \bibinfo {author}
  {\bibfnamefont {Z.-H.}\ \bibnamefont {Ci}}, \bibinfo {author} {\bibfnamefont
  {J.}~\bibnamefont {Gu}}, \bibinfo {author} {\bibfnamefont {T.}~\bibnamefont
  {Gao}}, \bibinfo {author} {\bibfnamefont {Q.-W.}\ \bibnamefont {Wang}}, \
  and\ \bibinfo {author} {\bibfnamefont {Z.-Y.}\ \bibnamefont {Shen}},\ }\href
  {\doibase 10.1364/OE.401829} {\bibfield  {journal} {\bibinfo  {journal} {Opt.
  Express}\ }\textbf {\bibinfo {volume} {28}},\ \bibinfo {pages} {29479}
  (\bibinfo {year} {2020})}\BibitemShut {NoStop}%
\bibitem [{\citenamefont {Scalcon}\ \emph {et~al.}(2022)\citenamefont
  {Scalcon}, \citenamefont {Agnesi}, \citenamefont {Avesani}, \citenamefont
  {Calderaro}, \citenamefont {Foletto}, \citenamefont {Stanco}, \citenamefont
  {Vallone},\ and\ \citenamefont {Villoresi}}]{2022Scalon}%
  \BibitemOpen
  \bibfield  {author} {\bibinfo {author} {\bibfnamefont {D.}~\bibnamefont
  {Scalcon}}, \bibinfo {author} {\bibfnamefont {C.}~\bibnamefont {Agnesi}},
  \bibinfo {author} {\bibfnamefont {M.}~\bibnamefont {Avesani}}, \bibinfo
  {author} {\bibfnamefont {L.}~\bibnamefont {Calderaro}}, \bibinfo {author}
  {\bibfnamefont {G.}~\bibnamefont {Foletto}}, \bibinfo {author} {\bibfnamefont
  {A.}~\bibnamefont {Stanco}}, \bibinfo {author} {\bibfnamefont
  {G.}~\bibnamefont {Vallone}}, \ and\ \bibinfo {author} {\bibfnamefont
  {P.}~\bibnamefont {Villoresi}},\ }\href {\doibase
  https://doi.org/10.1002/qute.202200051} {\bibfield  {journal} {\bibinfo
  {journal} {Advanced Quantum Technologies}\ }\textbf {\bibinfo {volume} {5}},\
  \bibinfo {pages} {2200051} (\bibinfo {year} {2022})}\BibitemShut {NoStop}%
\bibitem [{\citenamefont {Tang}\ \emph {et~al.}(2023)\citenamefont {Tang},
  \citenamefont {Zhou}, \citenamefont {Li}, \citenamefont {Xie}, \citenamefont
  {Xu}, \citenamefont {Sun}, \citenamefont {Zhang}, \citenamefont {Jiang},
  \citenamefont {Wang}, \citenamefont {Liu}, \citenamefont {Wu}, \citenamefont
  {Ma}, \citenamefont {Zheng}, \citenamefont {Jiang}, \citenamefont {Wang},
  \citenamefont {Zhao}, \citenamefont {Ma}, \citenamefont {Zhang},
  \citenamefont {Zhao}, \citenamefont {Bao},\ and\ \citenamefont
  {Tang}}]{2023Tang}%
  \BibitemOpen
  \bibfield  {author} {\bibinfo {author} {\bibfnamefont {Y.-L.}\ \bibnamefont
  {Tang}}, \bibinfo {author} {\bibfnamefont {C.}~\bibnamefont {Zhou}}, \bibinfo
  {author} {\bibfnamefont {D.-D.}\ \bibnamefont {Li}}, \bibinfo {author}
  {\bibfnamefont {Z.-L.}\ \bibnamefont {Xie}}, \bibinfo {author} {\bibfnamefont
  {M.-L.}\ \bibnamefont {Xu}}, \bibinfo {author} {\bibfnamefont
  {J.}~\bibnamefont {Sun}}, \bibinfo {author} {\bibfnamefont {Z.-X.}\
  \bibnamefont {Zhang}}, \bibinfo {author} {\bibfnamefont {L.-J.}\ \bibnamefont
  {Jiang}}, \bibinfo {author} {\bibfnamefont {L.-W.}\ \bibnamefont {Wang}},
  \bibinfo {author} {\bibfnamefont {G.-Q.}\ \bibnamefont {Liu}}, \bibinfo
  {author} {\bibfnamefont {K.}~\bibnamefont {Wu}}, \bibinfo {author}
  {\bibfnamefont {Y.}~\bibnamefont {Ma}}, \bibinfo {author} {\bibfnamefont
  {B.-R.}\ \bibnamefont {Zheng}}, \bibinfo {author} {\bibfnamefont {M.-S.}\
  \bibnamefont {Jiang}}, \bibinfo {author} {\bibfnamefont {Y.}~\bibnamefont
  {Wang}}, \bibinfo {author} {\bibfnamefont {Y.-K.}\ \bibnamefont {Zhao}},
  \bibinfo {author} {\bibfnamefont {Q.-L.}\ \bibnamefont {Ma}}, \bibinfo
  {author} {\bibfnamefont {D.}~\bibnamefont {Zhang}}, \bibinfo {author}
  {\bibfnamefont {M.-S.}\ \bibnamefont {Zhao}}, \bibinfo {author}
  {\bibfnamefont {W.-S.}\ \bibnamefont {Bao}}, \ and\ \bibinfo {author}
  {\bibfnamefont {S.-B.}\ \bibnamefont {Tang}},\ }\href {\doibase
  10.1364/OE.496723} {\bibfield  {journal} {\bibinfo  {journal} {Opt. Express}\
  }\textbf {\bibinfo {volume} {31}},\ \bibinfo {pages} {26335} (\bibinfo {year}
  {2023})}\BibitemShut {NoStop}%
\bibitem [{\citenamefont {Gr\"unenfelder}\ \emph {et~al.}(2023)\citenamefont
  {Gr\"unenfelder}, \citenamefont {Boaron}, \citenamefont {Resta},
  \citenamefont {Perrenoud}, \citenamefont {Rusca}, \citenamefont {Barreiro},
  \citenamefont {Houlmann}, \citenamefont {Sax}, \citenamefont {Stasi},
  \citenamefont {El-Khoury}, \citenamefont {H\"anggi}, \citenamefont
  {Bosshard}, \citenamefont {Bussi\`{e}res},\ and\ \citenamefont
  {Zbinden}}]{2023Grunenfelder}%
  \BibitemOpen
  \bibfield  {author} {\bibinfo {author} {\bibfnamefont {F.}~\bibnamefont
  {Gr\"unenfelder}}, \bibinfo {author} {\bibfnamefont {A.}~\bibnamefont
  {Boaron}}, \bibinfo {author} {\bibfnamefont {G.~V.}\ \bibnamefont {Resta}},
  \bibinfo {author} {\bibfnamefont {M.}~\bibnamefont {Perrenoud}}, \bibinfo
  {author} {\bibfnamefont {D.}~\bibnamefont {Rusca}}, \bibinfo {author}
  {\bibfnamefont {C.}~\bibnamefont {Barreiro}}, \bibinfo {author}
  {\bibfnamefont {R.}~\bibnamefont {Houlmann}}, \bibinfo {author}
  {\bibfnamefont {R.}~\bibnamefont {Sax}}, \bibinfo {author} {\bibfnamefont
  {L.}~\bibnamefont {Stasi}}, \bibinfo {author} {\bibfnamefont
  {S.}~\bibnamefont {El-Khoury}}, \bibinfo {author} {\bibfnamefont
  {E.}~\bibnamefont {H\"anggi}}, \bibinfo {author} {\bibfnamefont
  {N.}~\bibnamefont {Bosshard}}, \bibinfo {author} {\bibfnamefont
  {F.}~\bibnamefont {Bussi\`{e}res}}, \ and\ \bibinfo {author} {\bibfnamefont
  {H.}~\bibnamefont {Zbinden}},\ }\href {\doibase 10.1038/s41566-023-01168-2}
  {\bibfield  {journal} {\bibinfo  {journal} {Nature Photonics}\ }\textbf
  {\bibinfo {volume} {17}},\ \bibinfo {pages} {422} (\bibinfo {year}
  {2023})}\BibitemShut {NoStop}%
\bibitem [{\citenamefont {Francesconi}\ \emph {et~al.}(2024)\citenamefont
  {Francesconi}, \citenamefont {De~Lazzari}, \citenamefont {Ribezzo},
  \citenamefont {Vagniluca}, \citenamefont {Biagi}, \citenamefont {Occhipinti},
  \citenamefont {Zavatta},\ and\ \citenamefont {Bacco}}]{2024Francesconi}%
  \BibitemOpen
  \bibfield  {author} {\bibinfo {author} {\bibfnamefont {S.}~\bibnamefont
  {Francesconi}}, \bibinfo {author} {\bibfnamefont {C.}~\bibnamefont
  {De~Lazzari}}, \bibinfo {author} {\bibfnamefont {D.}~\bibnamefont {Ribezzo}},
  \bibinfo {author} {\bibfnamefont {I.}~\bibnamefont {Vagniluca}}, \bibinfo
  {author} {\bibfnamefont {N.}~\bibnamefont {Biagi}}, \bibinfo {author}
  {\bibfnamefont {T.}~\bibnamefont {Occhipinti}}, \bibinfo {author}
  {\bibfnamefont {A.}~\bibnamefont {Zavatta}}, \ and\ \bibinfo {author}
  {\bibfnamefont {D.}~\bibnamefont {Bacco}},\ }\href {\doibase
  https://doi.org/10.1002/qute.202300224} {\bibfield  {journal} {\bibinfo
  {journal} {Advanced Quantum Technologies}\ }\textbf {\bibinfo {volume} {7}},\
  \bibinfo {pages} {2300224} (\bibinfo {year} {2024})}\BibitemShut {NoStop}%
\bibitem [{\citenamefont {Kamin}\ and\ \citenamefont
  {L\"utkenhaus}(2024)}]{2024Kamin}%
  \BibitemOpen
  \bibfield  {author} {\bibinfo {author} {\bibfnamefont {L.}~\bibnamefont
  {Kamin}}\ and\ \bibinfo {author} {\bibfnamefont {N.}~\bibnamefont
  {L\"utkenhaus}},\ }\href {\doibase 10.1103/PhysRevResearch.6.043223}
  {\bibfield  {journal} {\bibinfo  {journal} {Phys. Rev. Res.}\ }\textbf
  {\bibinfo {volume} {6}},\ \bibinfo {pages} {043223} (\bibinfo {year}
  {2024})}\BibitemShut {NoStop}%
\bibitem [{\citenamefont {Tupkary}\ \emph {et~al.}(2025)\citenamefont
  {Tupkary}, \citenamefont {Tan}, \citenamefont {Nahar}, \citenamefont
  {Kamin},\ and\ \citenamefont {L\"utkenhaus}}]{2025Tupkary}%
  \BibitemOpen
  \bibfield  {author} {\bibinfo {author} {\bibfnamefont {D.}~\bibnamefont
  {Tupkary}}, \bibinfo {author} {\bibfnamefont {E.~Y.~Z.}\ \bibnamefont {Tan}},
  \bibinfo {author} {\bibfnamefont {S.}~\bibnamefont {Nahar}}, \bibinfo
  {author} {\bibfnamefont {L.}~\bibnamefont {Kamin}}, \ and\ \bibinfo {author}
  {\bibfnamefont {N.}~\bibnamefont {L\"utkenhaus}},\ }\href
  {https://arxiv.org/abs/2502.10340} {\bibfield  {journal} {\bibinfo  {journal}
  {\rm arXiv:2502.10340}\ } (\bibinfo {year} {2025})}\BibitemShut {NoStop}%
\bibitem [{\citenamefont {Zhang}\ \emph {et~al.}(2021)\citenamefont {Zhang},
  \citenamefont {Coles}, \citenamefont {Winick}, \citenamefont {Lin},\ and\
  \citenamefont {L\"utkenhaus}}]{2021Zhang}%
  \BibitemOpen
  \bibfield  {author} {\bibinfo {author} {\bibfnamefont {Y.}~\bibnamefont
  {Zhang}}, \bibinfo {author} {\bibfnamefont {P.~J.}\ \bibnamefont {Coles}},
  \bibinfo {author} {\bibfnamefont {A.}~\bibnamefont {Winick}}, \bibinfo
  {author} {\bibfnamefont {J.}~\bibnamefont {Lin}}, \ and\ \bibinfo {author}
  {\bibfnamefont {N.}~\bibnamefont {L\"utkenhaus}},\ }\href {\doibase
  10.1103/PhysRevResearch.3.013076} {\bibfield  {journal} {\bibinfo  {journal}
  {Phys. Rev. Res.}\ }\textbf {\bibinfo {volume} {3}},\ \bibinfo {pages}
  {013076} (\bibinfo {year} {2021})}\BibitemShut {NoStop}%
\bibitem [{\citenamefont {Kamin}\ \emph
  {et~al.}(2025{\natexlab{a}})\citenamefont {Kamin}, \citenamefont {Tupkary},\
  and\ \citenamefont {L\"utkenhaus}}]{2025KaminPost}%
  \BibitemOpen
  \bibfield  {author} {\bibinfo {author} {\bibfnamefont {L.}~\bibnamefont
  {Kamin}}, \bibinfo {author} {\bibfnamefont {D.}~\bibnamefont {Tupkary}}, \
  and\ \bibinfo {author} {\bibfnamefont {N.}~\bibnamefont {L\"utkenhaus}},\
  }\href {https://arxiv.org/abs/2502.05382} {\bibfield  {journal} {\bibinfo
  {journal} {\rm arXiv:2502.05382}\ } (\bibinfo {year}
  {2025}{\natexlab{a}})}\BibitemShut {NoStop}%
\bibitem [{\citenamefont {Kamin}\ \emph
  {et~al.}(2025{\natexlab{b}})\citenamefont {Kamin}, \citenamefont
  {Burniston},\ and\ \citenamefont {Tan}}]{2025KaminRenyi}%
  \BibitemOpen
  \bibfield  {author} {\bibinfo {author} {\bibfnamefont {L.}~\bibnamefont
  {Kamin}}, \bibinfo {author} {\bibfnamefont {J.}~\bibnamefont {Burniston}}, \
  and\ \bibinfo {author} {\bibfnamefont {E.~Y.~Z.}\ \bibnamefont {Tan}},\
  }\href {https://arxiv.org/abs/2504.12248} {\bibfield  {journal} {\bibinfo
  {journal} {\rm arXiv:2504.12248}\ } (\bibinfo {year}
  {2025}{\natexlab{b}})}\BibitemShut {NoStop}%
\bibitem [{\citenamefont {Brassard}\ \emph {et~al.}(2000)\citenamefont
  {Brassard}, \citenamefont {L\"{u}tkenhaus}, \citenamefont {Mor},\ and\
  \citenamefont {Sanders}}]{2000Brassard}%
  \BibitemOpen
  \bibfield  {author} {\bibinfo {author} {\bibfnamefont {G.}~\bibnamefont
  {Brassard}}, \bibinfo {author} {\bibfnamefont {N.}~\bibnamefont
  {L\"{u}tkenhaus}}, \bibinfo {author} {\bibfnamefont {T.}~\bibnamefont {Mor}},
  \ and\ \bibinfo {author} {\bibfnamefont {B.}~\bibnamefont {Sanders}},\ }\href
  {https://doi.org/10.1103/PhysRevLett.85.1330} {\bibfield  {journal} {\bibinfo
   {journal} {Phy. Rev. Lett.}\ }\textbf {\bibinfo {volume} {85}},\ \bibinfo
  {pages} {1330} (\bibinfo {year} {2000})}\BibitemShut {NoStop}%
\bibitem [{\citenamefont {Ma}\ \emph {et~al.}(2005)\citenamefont {Ma},
  \citenamefont {Qi}, \citenamefont {Zhao},\ and\ \citenamefont {Lo}}]{2005Ma}%
  \BibitemOpen
  \bibfield  {author} {\bibinfo {author} {\bibfnamefont {X.}~\bibnamefont
  {Ma}}, \bibinfo {author} {\bibfnamefont {B.}~\bibnamefont {Qi}}, \bibinfo
  {author} {\bibfnamefont {Y.}~\bibnamefont {Zhao}}, \ and\ \bibinfo {author}
  {\bibfnamefont {H.-K.}\ \bibnamefont {Lo}},\ }\href {\doibase
  10.1103/PhysRevA.72.012326} {\bibfield  {journal} {\bibinfo  {journal} {Phys.
  Rev. A}\ }\textbf {\bibinfo {volume} {72}},\ \bibinfo {pages} {012326}
  (\bibinfo {year} {2005})}\BibitemShut {NoStop}%
\bibitem [{\citenamefont {Sanari}\ \emph {et~al.}(2024)\citenamefont {Sanari},
  \citenamefont {Taniguchi}, \citenamefont {Miura}, \citenamefont {Takahashi},
  \citenamefont {Takasugi}, \citenamefont {Lo}, \citenamefont {Ikuta},
  \citenamefont {Honjo},\ and\ \citenamefont {Takesue}}]{2024Sanari}%
  \BibitemOpen
  \bibfield  {author} {\bibinfo {author} {\bibfnamefont {Y.}~\bibnamefont
  {Sanari}}, \bibinfo {author} {\bibfnamefont {A.}~\bibnamefont {Taniguchi}},
  \bibinfo {author} {\bibfnamefont {M.}~\bibnamefont {Miura}}, \bibinfo
  {author} {\bibfnamefont {H.}~\bibnamefont {Takahashi}}, \bibinfo {author}
  {\bibfnamefont {K.}~\bibnamefont {Takasugi}}, \bibinfo {author}
  {\bibfnamefont {H.-P.}\ \bibnamefont {Lo}}, \bibinfo {author} {\bibfnamefont
  {T.}~\bibnamefont {Ikuta}}, \bibinfo {author} {\bibfnamefont
  {T.}~\bibnamefont {Honjo}}, \ and\ \bibinfo {author} {\bibfnamefont
  {H.}~\bibnamefont {Takesue}},\ }in\ \href
  {https://opg.optica.org/abstract.cfm?URI=CLEOPR-2024-Fr1C_2} {\emph {\bibinfo
  {booktitle} {2024 Conference on Lasers and Electro-Optics Pacific Rim
  (CLEO-PR)}}}\ (\bibinfo  {publisher} {Optica Publishing Group},\ \bibinfo
  {year} {2024})\BibitemShut {NoStop}%
\bibitem [{\citenamefont {Luo}\ \emph {et~al.}(2024)\citenamefont {Luo},
  \citenamefont {Cheng}, \citenamefont {Mao},\ and\ \citenamefont
  {Li}}]{2024Luo}%
  \BibitemOpen
  \bibfield  {author} {\bibinfo {author} {\bibfnamefont {Y.}~\bibnamefont
  {Luo}}, \bibinfo {author} {\bibfnamefont {X.}~\bibnamefont {Cheng}}, \bibinfo
  {author} {\bibfnamefont {H.-K.}\ \bibnamefont {Mao}}, \ and\ \bibinfo
  {author} {\bibfnamefont {Q.}~\bibnamefont {Li}},\ }\href {\doibase
  10.3390/math12142243} {\bibfield  {journal} {\bibinfo  {journal}
  {Mathematics}\ }\textbf {\bibinfo {volume} {12}} (\bibinfo {year} {2024}),\
  10.3390/math12142243}\BibitemShut {NoStop}%
\bibitem [{\citenamefont {Li}\ \emph {et~al.}(2022)\citenamefont {Li},
  \citenamefont {Niu}, \citenamefont {Feng},\ and\ \citenamefont
  {Chen}}]{2022Li}%
  \BibitemOpen
  \bibfield  {author} {\bibinfo {author} {\bibfnamefont {B.}~\bibnamefont
  {Li}}, \bibinfo {author} {\bibfnamefont {Y.}~\bibnamefont {Niu}}, \bibinfo
  {author} {\bibfnamefont {Y.}~\bibnamefont {Feng}}, \ and\ \bibinfo {author}
  {\bibfnamefont {X.}~\bibnamefont {Chen}},\ }\href
  {https://doi.org/10.1007/s11801-022-2036-3} {\bibfield  {journal} {\bibinfo
  {journal} {Optoelectronics Letters}\ }\textbf {\bibinfo {volume} {18}},\
  \bibinfo {pages} {647} (\bibinfo {year} {2022})}\BibitemShut {NoStop}%
\bibitem [{\citenamefont {Kato}(2020)}]{2020Kato}%
  \BibitemOpen
  \bibfield  {author} {\bibinfo {author} {\bibfnamefont {G.}~\bibnamefont
  {Kato}},\ }\href {https://arxiv.org/abs/2002.04357} {\bibfield  {journal}
  {\bibinfo  {journal} {\rm arXiv:2002.04357}\ } (\bibinfo {year}
  {2020})}\BibitemShut {NoStop}%
\bibitem [{\citenamefont {Wang}\ \emph {et~al.}(2023)\citenamefont {Wang},
  \citenamefont {Wang}, \citenamefont {Hu}, \citenamefont {Zapatero},
  \citenamefont {Qian}, \citenamefont {Qi}, \citenamefont {Curty},\ and\
  \citenamefont {Lo}}]{2023Wang}%
  \BibitemOpen
  \bibfield  {author} {\bibinfo {author} {\bibfnamefont {W.}~\bibnamefont
  {Wang}}, \bibinfo {author} {\bibfnamefont {R.}~\bibnamefont {Wang}}, \bibinfo
  {author} {\bibfnamefont {C.}~\bibnamefont {Hu}}, \bibinfo {author}
  {\bibfnamefont {V.}~\bibnamefont {Zapatero}}, \bibinfo {author}
  {\bibfnamefont {L.}~\bibnamefont {Qian}}, \bibinfo {author} {\bibfnamefont
  {B.}~\bibnamefont {Qi}}, \bibinfo {author} {\bibfnamefont {M.}~\bibnamefont
  {Curty}}, \ and\ \bibinfo {author} {\bibfnamefont {H.-K.}\ \bibnamefont
  {Lo}},\ }\href {\doibase 10.1103/PhysRevLett.130.220801} {\bibfield
  {journal} {\bibinfo  {journal} {Phys. Rev. Lett.}\ }\textbf {\bibinfo
  {volume} {130}},\ \bibinfo {pages} {220801} (\bibinfo {year}
  {2023})}\BibitemShut {NoStop}%
\bibitem [{\citenamefont {Zapatero}\ \emph {et~al.}(2023)\citenamefont
  {Zapatero}, \citenamefont {Wang},\ and\ \citenamefont
  {Curty}}]{2023Zapatero}%
  \BibitemOpen
  \bibfield  {author} {\bibinfo {author} {\bibfnamefont {V.}~\bibnamefont
  {Zapatero}}, \bibinfo {author} {\bibfnamefont {W.}~\bibnamefont {Wang}}, \
  and\ \bibinfo {author} {\bibfnamefont {M.}~\bibnamefont {Curty}},\ }\href
  {\doibase 10.1088/2058-9565/acbc46} {\bibfield  {journal} {\bibinfo
  {journal} {Quantum Science and Technology}\ }\textbf {\bibinfo {volume}
  {8}},\ \bibinfo {pages} {025014} (\bibinfo {year} {2023})}\BibitemShut
  {NoStop}%
\bibitem [{\citenamefont {Zapatero}\ and\ \citenamefont
  {Curty}(2024)}]{2024Zapatero}%
  \BibitemOpen
  \bibfield  {author} {\bibinfo {author} {\bibfnamefont {V.}~\bibnamefont
  {Zapatero}}\ and\ \bibinfo {author} {\bibfnamefont {M.}~\bibnamefont
  {Curty}},\ }\href {\doibase 10.1103/PhysRevApplied.21.014018} {\bibfield
  {journal} {\bibinfo  {journal} {Phys. Rev. Appl.}\ }\textbf {\bibinfo
  {volume} {21}},\ \bibinfo {pages} {014018} (\bibinfo {year}
  {2024})}\BibitemShut {NoStop}%
\bibitem [{\citenamefont {Wang}\ \emph {et~al.}(2025)\citenamefont {Wang},
  \citenamefont {Tupkary},\ and\ \citenamefont {Nahar}}]{2025Wang}%
  \BibitemOpen
  \bibfield  {author} {\bibinfo {author} {\bibfnamefont {Z.}~\bibnamefont
  {Wang}}, \bibinfo {author} {\bibfnamefont {D.}~\bibnamefont {Tupkary}}, \
  and\ \bibinfo {author} {\bibfnamefont {S.}~\bibnamefont {Nahar}},\ }\href
  {https://arxiv.org/abs/2508.21486} {\bibfield  {journal} {\bibinfo  {journal}
  {\rm arXiv:2508.21486}\ } (\bibinfo {year} {2025})}\BibitemShut {NoStop}%
\end{thebibliography}%
\bibliographystyle{apsrev4-1}

\end{document}